\documentclass[aps,pre,twocolumn,superscriptaddress,longbibliography,nofootinbib,showkeys,showpacs]{revtex4-1}

\usepackage{amsfonts}
\usepackage{amsmath}
\usepackage{appendix}
\usepackage{bm}
\usepackage{braket}
\usepackage{caption}
\usepackage{xcolor}
\usepackage{comment}
\usepackage{enumerate}
\usepackage[top=2cm,left=2cm,right=2cm,bottom=2cm]{geometry}
\usepackage{graphicx}
\usepackage{hyperref}
\usepackage[utf8]{inputenc}
\usepackage{mathptmx}

\begin{document}

\colorlet{CV}{.}

\title{Quantum annealing with trigger Hamiltonians: Application to 2-satisfiability and nonstoquastic problems}

\author{Vrinda Mehta}
\affiliation{Institute for Advanced Simulation, J\"ulich Supercomputing Centre,\\
Forschungszentrum J\"ulich, D-52425 J\"ulich, Germany.}
\affiliation{RWTH Aachen University, D-52056 Aachen, Germany.}

\author{Fengping Jin}
\affiliation{Institute for Advanced Simulation, J\"ulich Supercomputing Centre,\\
Forschungszentrum J\"ulich, D-52425 J\"ulich, Germany.}

\author{Hans De Raedt}
\affiliation{Institute for Advanced Simulation, J\"ulich Supercomputing Centre,\\
Forschungszentrum J\"ulich, D-52425 J\"ulich, Germany.}
\affiliation{Zernike Institute for Advanced Materials,\\
University of Groningen, Nijenborgh 4, NL-9747 AG Groningen, The Netherlands.}

\author{Kristel Michielsen}
\email{k.michielsen@fz-juelich.de}
\thanks{Corresponding author}
\affiliation{Institute for Advanced Simulation, J\"ulich Supercomputing Centre,\\
Forschungszentrum J\"ulich, D-52425 J\"ulich, Germany.}
\affiliation{RWTH Aachen University, D-52056 Aachen, Germany.}
\affiliation{JARA-CSD, Jülich-Aachen Research Alliance, 52425 Jülich, Germany.}

\date{\today}

\begin{abstract}
We study the performance of quantum annealing for two sets of problems, namely, 2-satisfiability (2-SAT) problems represented by Ising-type Hamiltonians, and nonstoquastic problems which are obtained by adding extra couplings to the 2-SAT problem Hamiltonians. In addition, we add to the transverse Ising-type Hamiltonian used for quantum annealing a third term, the trigger Hamiltonian with ferromagnetic or antiferromagnetic couplings, which vanishes at the beginning and end of the annealing process. We also analyze some problem instances using  the energy spectrum, average energy or overlap of the state during the evolution with the instantaneous low lying eigenstates of the Hamiltonian, and  identify some nonadiabatic mechanisms which can enhance the performance of quantum annealing.
\end{abstract}

\maketitle

\section{Introduction}

Applications of optimization problems are very diverse and 
\textcolor{CV}{include}, among others, scheduling chemotherapy for cancer patients \cite{zietz1979mathematical,shi2014survey} and aircraft assignment problems \cite{tail,vikstaal2020applying}. Modeling these problems to fit real-life applications requires a large number of variables. Many of these problems are NP-hard, implying that the time required to solve them grows exponentially with their size \cite{garey1979computers}. Various physical and computational techniques have been, therefore, devised and utilized to obtain the required solution.

One such physical technique is that of simulated annealing, which requires the problem to be mapped as a classical Hamiltonian, whose ground state then encodes the solution to the optimization problem \cite{kirkpatrick1983optimization}. Adding thermal fluctuations to the system keeps it from getting trapped in a local minimum. However, if the energy landscape of the Hamiltonian becomes too involved, \textcolor{CV}{i.e., the global minimum becomes surrounded by many local minima,} with limited computing resources, simulated annealing can fail to yield the correct ground state of the Hamiltonian. It was in this spirit that the method of quantum annealing was first employed to solve an optimization problem, aiming to utilize quantum tunneling to obtain the required solution \cite{apolloni1989quantum,kadowaki1998quantum,farhi2000quantum,farhi2001quantum,albash2018adiabatic}.

For quantum annealing, the minimum energy gap that occurs during the annealing process between the ground state and the first-excited state of the Hamiltonian, is of great importance\textcolor{CV}{, in particular in the adiabatic limit of quantum annealing}. However, it has been found that for the Ising Hamiltonian, which is used as a standard model Hamiltonian for encoding optimization problems, the minimum energy gap may be accompanied by a quantum phase transition, and for difficult problems, it can close exponentially fast as the system size grows \cite{young2010first,neuhaus2011classical,knysh2016zero,hauke2020perspectives}. Therefore, the time required to ascertain an adiabatic evolution of the state also increases exponentially for these problems. Thus to utilize quantum annealing well, it is advantageous to find ways to enlarge the minimum energy gap. In this work we attempt to do this by introducing a trigger Hamiltonian to the existing D-Wave type transverse Ising-spin Hamiltonian \cite{farhi2002quantum,zeng2016schedule,hormozi2017nonstoquastic,steering2018,albash2018adiabatic}.

We study two types of trigger Hamiltonians: the ferromagnetic trigger Hamiltonian and antiferromagnetic trigger Hamiltonian. While the addition of the former still keeps the whole Hamiltonian stoquastic (i.e., it has real and nonpositive off-diagonal matrix elements in the computational basis), the latter makes the Hamiltonian nonstoquastic. It has been suggested that nonstoquastic Hamiltonians can be more advantageous for the efficiency of quantum annealing \cite{seki2012quantum,seki2015quantum,hormozi2017nonstoquastic,Nishimori_2017,crosson2020signing}.
Lastly, we also examine the performance of quantum annealing for nonstoquastic problems, in contrast with the Ising problems which are stoquastic (have only diagonal matrix elements).

In Sec.~\ref{theory}, we briefly discuss the basics of quantum annealing, and introduce the trigger Hamiltonian. Section~\ref{problems} describes the problem sets for which the quantum annealing algorithm will be performed. Sections~\ref{results_2sat} and \ref{results_nonstoq} showcase the results and analysis. We finally conclude our observations in Sec.~\ref{conclusion}.

\section{Quantum annealing and trigger Hamiltonians}
\label{theory}

The quantum annealing algorithm requires one to start in the ground state of an initial Hamiltonian, $H_I$, whose ground state is easy to determine and prepare. With the help of a control parameter, $s$, defined as $t/T_A$, where $T_A$ is the total annealing time and $t$ denotes the time, the time-dependent Hamiltonian is swept towards the problem Hamiltonian, $H_P$, whose ground state encodes the solution to the problem at hand. The time-dependent Hamiltonian, $H(t)$, thus has the following form:
\begin{equation}
    H(t) = A(s(t)) H_I + B(s(t)) H_P,
    \label{eq_annealing}
\end{equation}
where functions $A$ and $B$ control the sweeping scheme, such that $A(0) \approx 1$, and $B(0) \approx 0$, while $A(1) \approx 0$, and $B(1) \approx 1$.\\
The problem of finding the optimal solution, therefore reduces to solving the time-dependent Schr\"{o}dinger equation for the resulting $H(t)$,
\begin{equation}
    i \frac{\partial}{\partial t} \ket{\psi} = H(t) \ket{\psi}\textcolor{CV}{,}
\end{equation}
\textcolor{CV}{where $\hbar$ is set to 1, making the subsequent quantities involved in this paper dimensionless.}\\
The problem Hamiltonian is generally chosen to be of Ising type, i.e.,
\begin{equation}
    H_P = -\sum_{i} {h_i}^z{\sigma_i}^z - \sum_{i,j} {J_{i,j}^z} {\sigma_i}^z {\sigma_j}^z,
    \label{eq_hp}
\end{equation}
while the initial Hamiltonian is chosen to be
\begin{equation}
    H_I = -\sum_{i} {h_i}^x{\sigma_i}^x,
\end{equation}
where ${\sigma_i}^x$ and ${\sigma_i}^z$ \textcolor{CV}{are the Pauli matrices} acting on the $i$th spin. Here, $J_{i,j}^z$ represents the $z$ coupling between the $i$th and $j$th spins, and $h_i^{x(z)}$ represents the $x(z)$ magnetic field acting on the $i$th spin, whereby $h_i^x$ is generally chosen to be 1 for all the spins. The ground state of the initial Hamiltonian is the uniform superposition state.\\

The quantum adiabatic theorem assures that if the sweeping from the initial Hamiltonian to the final Hamiltonian is done slowly enough (as determined by the minimum energy gap between the ground state and the first-excited state of the Hamiltonian), the state of the system shall remain close to the ground state of the instantaneous Hamiltonian, if one starts in the ground state of the initial Hamiltonian \cite{kato1950adiabatic,born1928beweis,jansen2007bounds,amin2009consistency}. Mathematically, for the adiabatic theorem to hold, the annealing time $T_A$ should satisfy
\begin{equation}
    T_A \gg {\max_{0 \leq s \leq 1}} \frac{||\bra{1(s)} \frac{dH}{ds} \ket{0(s)}||}{\Delta (s)^2}, 
    \label{eq:adiabatic}
\end{equation}
where $\ket{0(s)}$ and $\ket{1(s)}$ represent the ground state and the first-excited state of the instantaneous Hamiltonian respectively, and $\Delta(s)$ is the energy gap between these states.\\
Thus, the size and shape of the minimum energy gap that is encountered during the evolution, is crucial for determining the difficulty (in terms of the annealing time required to ensure that the evolution is adiabatic, or, equivalently, the probability for the evolution to stay adiabatic for a given annealing time) of the problem at hand. \\

In an attempt to modify this minimum energy gap, we add a third term, the trigger Hamiltonian, $H_T$, to Eq.~(\ref{eq_annealing}). The trigger term vanishes at the start and end of the annealing process to preserve the ground state of the initial and the final Hamiltonians. We then study the effects that this additional term might have on the energy spectrum of the Hamiltonian and the probability of the state staying in the ground state of the Hamiltonian during the evolution. The time-dependent Hamiltonian, upon choosing a linear scheme, thus becomes:
\begin{equation}
    H(t)= (1-s) H_I + s(1-s) H_T + s H_P.
    \label{eq:totalHamil}
\end{equation}
\textcolor{CV}{The couplings in the trigger Hamiltonian are chosen according to the graph of the problem Hamiltonian}, and thus
\begin{equation}
    H_T=- g \sum_{i,j} {J_{i,j}^x} {\sigma_i}^x {\sigma_j}^x,
\end{equation}
where the parameter $g$ controls the strength of the trigger Hamiltonian. For a two-spin system it is possible to obtain an analytical solution for the energy spectrum of the Hamiltonian.~(\ref{eq:totalHamil}), and is given in Appendix~\ref{sec:appendix_A}.\\

Furthermore, there are two types of trigger Hamiltonians chosen for this work, one with ferromagnetic transverse couplings (\textcolor{CV}{$J_{i,j}^x$} = 1), referred to as the ferromagnetic trigger, and the other with antiferromagnetic transverse couplings (\textcolor{CV}{$J_{i,j}^x$} = -1), referred to as the antiferromagnetic trigger.\\

\section{The optimization problems}
\label{problems}

A 2-SAT (satisfiability) problem is a Boolean expression in the form of a conjunction of $M$ clauses involving $N$ variables and their negations, where each clause is a disjunction of two literals. Thus the cost function has the following form:
\begin{equation}
    F = (L_{1,1} \lor L_{1,2}) \land (L_{2,1} \lor L_{2,2}) \land ....\land (L_{M,1} \lor L_{M,2}),
\end{equation}
where literal $L_{\alpha,j}$, with $\alpha=1,2,...,M$ and $j=1,2$, can be a variable $x_i$ or its negation $\overline{x_i}$. A problem is considered to be satisfiable if one can find an assignment to the variables such that the cost function $F$ is true. \\
These problems can be reformulated as Ising Hamiltonians \cite{neuhaus2014monte}, where the physical degree of freedom are the spins $s_i$, using the mapping
\begin{equation}
    \textcolor{CV}{C_{2SAT}} = \sum_{\alpha=1}^{M}  \textcolor{CV}{c_{2SAT}}(\epsilon_{\alpha,1} s_{i[\alpha,1]}, \epsilon_{\alpha,2}s_{i[\alpha,2]}), 
    \label{eq:mapping}
\end{equation}
where $i[\alpha,j]$ represents the variable $i$ involved as the $j$th term in the clause $\alpha$ of the cost function. If variable $x_i$ is used as this term,  $\epsilon_{\alpha,j}=1$, while for its negation $\overline{x_i}$, $\epsilon_{\alpha,j}=-1$. As an example, for \textcolor{CV}{clauses $(x_1 \lor x_2)$, $(\overline{x_1} \lor x_2)$, $(x_1 \lor \overline{x_2})$, and $(\overline{x_1} \lor \overline{x_2})$, we have $c_{2SAT}=s_1s_2-(s_1+s_2)+1$, $-s_1s_2-(-s_1+s_2)+1$, $-s_1s_2-(s_1-s_2)+1$, and $s_1s_2+(s_1+s_2)+1$, respectively}. \\
To utilize quantum annealing to find the truth assignment to the cost function, the classical spin $s_i$ is replaced by the quantum spin ${\sigma_i}^z$. \\

\textcolor{CV}{	The Hamiltonians corresponding to 2-SAT problems consist of 2-local coupling terms, as in Eq.~(\ref{eq_hp}), which makes them physically relevant in terms of potential implementations on real devices. Although the mapping given in Eq.~(\ref{eq:mapping}) can be extended to $K$-SAT problems ($K>2$), the resulting Hamiltonian has $K$-local coupling terms. There exist methods that can reduce the $K$-local couplings to 2-local couplings by introducing auxiliary variables, but this generally increases the degeneracy of the ground state. For making the mechanisms accompanied by the quantum evolution of the state of the system easily identifiable, we restrict ourselves to 2-SAT problems with a unique satisfying assignment for this work.\\
At this point, it should be noted that there are classical algorithms that can solve 2-SAT problems in polynomial time using a directed graph and identifying strongly connected components of the problem. On the other hand, these problems need to be mapped to the Ising model in order that quantum annealing can be used for solving them. Once mapped, quantum annealing cannot distinguish between the problem Hamiltonian from the mapping of a 2-SAT problem and a spin-glass Ising problem. Furthermore, previous studies have indicated an exponential scaling of quantum annealing for solving both 2-SAT and NP-Complete 3-SAT problems based on the (extended) mapping given in Eq.~(\ref{eq:mapping}) \cite{neuhaus2011classical,neuhaus2014quantum}. \\
We have chosen number of clauses, $M=N+1$, as the satisfiability threshold for a random set of 2-SAT problems, defined as the threshold at which the cost functions become unsatisfiable in the mean with a high probability, occurs approximately at $M/N \approx 1$ \cite{neuhaus2014monte}.\\
In this work we first consider two sets of 2-SAT problems, one with 12-variable problems and another with 18-variable problems, for which the problem Hamiltonian is of the form as given in Eq.~(\ref{eq_hp}). Both sets have 1000 different problems each. These problems have a unique known ground state (chosen using the brute force search method) and a highly degenerate first-excited state \cite{neuhaus2014monte}.
}

Finally, we extend the set of existing problems to nonstoquastic problems, by adding $y$ couplings \textcolor{CV}{with strength 0.5 according to the graph of the problem Hamiltonian}. Hence, the Hamiltonians of these problems read
\begin{equation}
    H_P = -\sum_{i} {h_i}^z{\sigma_i}^z - \sum_{i,j}\big( {J_{i,j}^z} {\sigma_i}^z {\sigma_j}^z + \frac{1}{2} {\sigma_i}^y {\sigma_j}^y\big).
\end{equation}
Also for the nonstoquastic problems we consider two sets, one with 12 variables and one with 18 variables. One major point of difference \textcolor{CV}{observed numerically in the properties of the nonstoquastic problems} is that their minimum energy gaps are approximately $O$(10) smaller than those of the 2-SAT problems. Additionally, some of the nonstoquastic problems have multiple (maximum of three) anticrossings between the ground state and the first-excited state. For these problems as well, we add the ferromagnetic and antiferromagnetic triggers to the annealing Hamiltonian~(\ref{eq_annealing}) to study the behavior that these sets might exhibit as a response to adding the trigger. 
 
\section{Quantum annealing for 2-SAT problems}
\label{results_2sat}

The first section discusses the results for the 2-SAT data obtained for the quantum annealing algorithm without using a trigger Hamiltonian. In the following sections, we address the effects of adding the two kinds of trigger Hamiltonians with trigger strengths $g=0.5,1.0,2.0$.\\
For implementing the time-dependent Schr\"{o}dinger equation to obtain the final state of the system at the end of the annealing process, we use the second-order Suzuki-Trotter product formula algorithm
\cite{trotter1959product,suzuki1977monte,de1987product,huyghebaert1990product,Hatano_2005}. We employ the Lanczos algorithm to determine the energy spectra and the minimum energy gaps of the problems \cite{cullum2002lanczos}. In Appendix~\ref{sec:appendix_B}, we give a review of the second-order Suzuki-Trotter product formula algorithm. We performed the simulation for systems with 12 variables using workstations equipped with Intel Core i7-8700 and 32 GB memory, and for larger systems, we used supercomputers JURECA and JUWELS of the J\"{u}lich Supercomputing Centre at Forschungszentrum J\"{u}lich \cite{jureca,JUWELS}.\\
Since the exact ground state of the considered problems is known, we can define the success probability as the \textcolor{CV}{squared} overlap of the final state obtained at the end of the annealing process, with the known ground state. For each problem, we determine the success probability for the annealing times $T_A=10,100,1000,10000$.

We \textcolor{CV}{mainly} show the results for the 18-variable 2-SAT problems. The results for the 12-variable problems are very similar and can be found in Ref.~\cite{Mehta:868380}.

\subsection{Without the trigger Hamiltonian}
\begin{figure}
    \centering
    \includegraphics[scale=0.7]{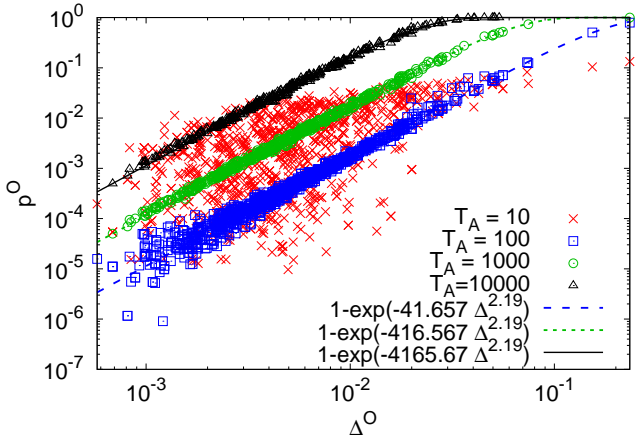}   
    \caption{(Color online) Success probability $p^O$ versus minimum energy gap $\Delta^O$ for 18-variable 2-SAT problems for various annealing times $T_A$. The Hamiltonian determining the annealing scheme does not contain any trigger Hamiltonian. The lines are fits to the data.}
    \label{fig:orig_succvsgap}
\end{figure}

We first analyze the performance of the quantum annealing algorithm for solving the 2-SAT problems without adding any trigger Hamiltonian to the Hamiltonian.~(\ref{eq_annealing}). Figure~\ref{fig:orig_succvsgap} shows the success probability $p^O$ as a function of the minimum energy gap $\Delta^O$ \textcolor{CV}{(where $\Delta^O=\min_{0 \leq s \leq 1} \Delta^O(s)$, and similarly for the following sections)}, for various $T_A$. We have fitted the data for $T_A=10000$ to the function $p=1-\exp({-a\Delta^{b}})$, and found the fitting parameters $a=4165.67$ and $b=2.19$. For $T_A=1000$ and $T_A=100$, $a$ is reduced by a factor 10 and 100 respectively, without changing $b$. The fitting curves match the data well, which implies that the system can be described by a Landau-Zener like model for which $b=2$, and the parameter $a$ is inversely proportional to the sweep speed $c$ \cite{landau1932theorie,zener1932non,de1997theory}. Contrary to the conditions under which the Landau-Zener formula is derived, the system at hand is not a two-level system, and the annealing process is carried over a finite amount of time. Nevertheless, the formula predicts the observations for long annealing times rather well, but with the parameter $b$ slightly deviating from two. In addition, as predicted by the Landau-Zener formula, the data show that an increase in the annealing time leads to an increase of the success probability.

From Fig.~\ref{fig:orig_succvsgap} it can be seen that, for $T_A=10$, the data show significant scattering, suggesting that the chosen annealing time is not long enough for the Landau-Zener formula to explain the success probabilities for a majority of the problems. \textcolor{CV}{As will be explained in more detail in ~Sec.~\ref{sec:antiferro}, in this case, the nonadiabatic mechanism of fast annealing is responsible for large success probabilities despite of the small minimum-energy gaps.} Because of this large scattering of the data for $T_A=10$, we have omitted fitting for this data. For an increasing annealing time, the scattering in the data reduces, and the system can be described by the Landau-Zener formula. 

\subsection{With the ferromagnetic trigger Hamiltonian}

\begin{figure}
    \centering
    \includegraphics[scale=0.7]{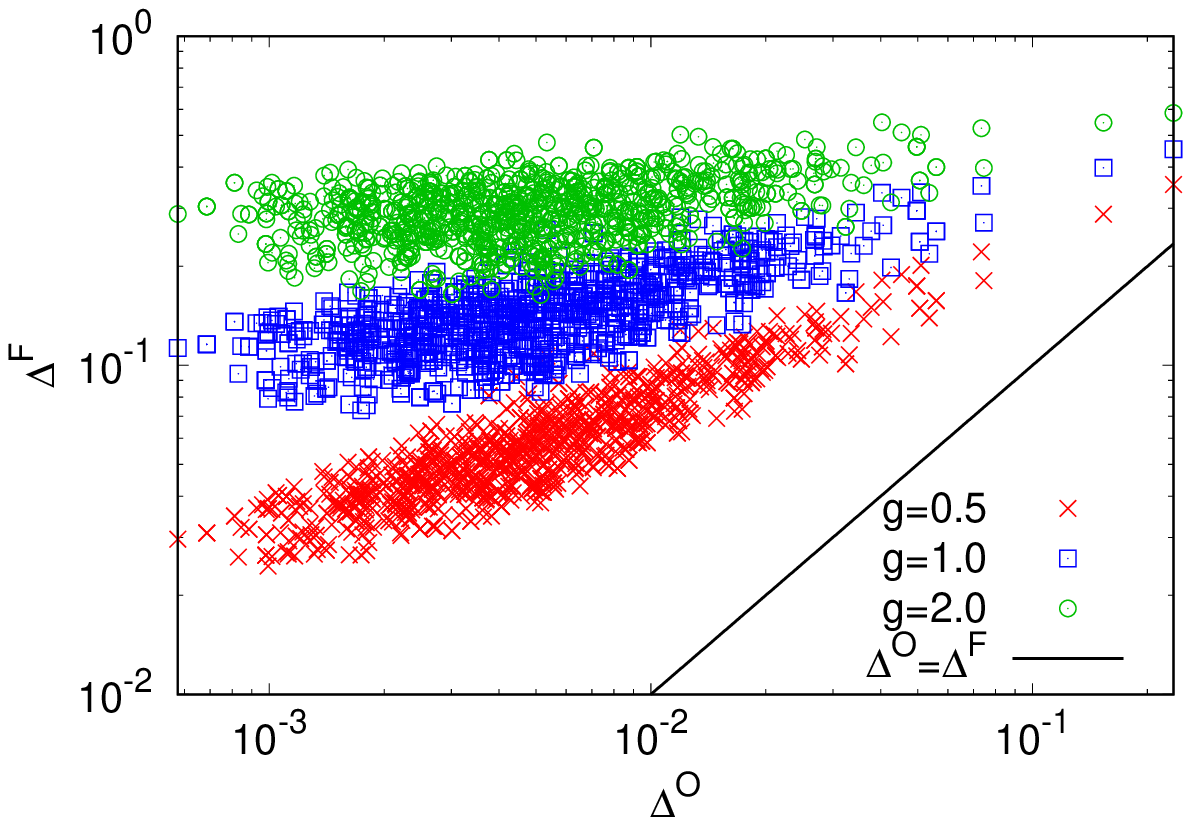}   
    \caption{(Color online) Minimum energy gap $\Delta^F$ for 18-variable 2-SAT problems after adding the ferromagnetic trigger Hamiltonian to the Hamiltonian~(\ref{eq_annealing}) versus minimum energy gap $\Delta^O$ without adding the trigger Hamiltonian, and this for various trigger strengths $g$.}
    \label{fig:ferro_mingap}
\end{figure}
Here, we discuss the results for the 2-SAT problems obtained for quantum annealing with the ferromagnetic trigger Hamiltonian added to the Hamiltonian~(\ref{eq_annealing}).

\begin{figure*}[ht]
    \centering
      \begin{minipage}[l]{0.32\textwidth}
         \includegraphics[scale=0.47]{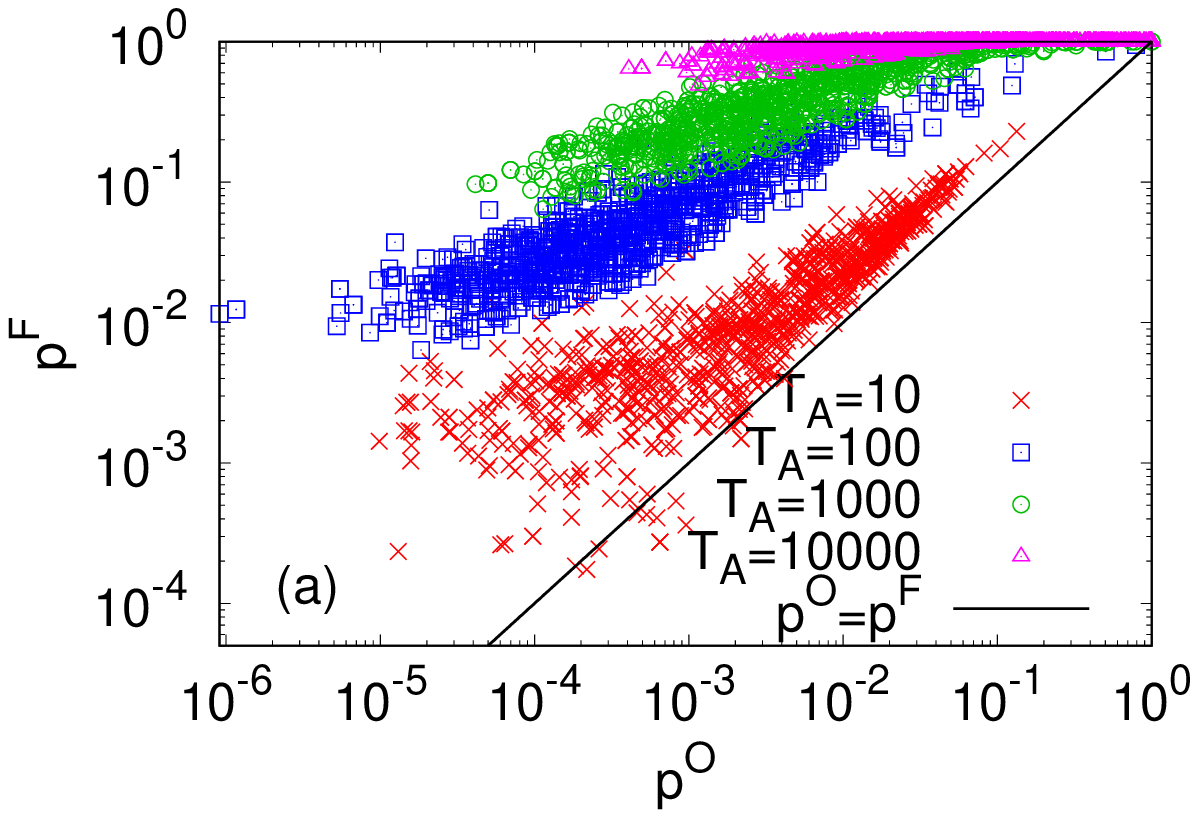}
     \end{minipage}
     \begin{minipage}[c]{0.32\textwidth}
         \includegraphics[scale=0.47]{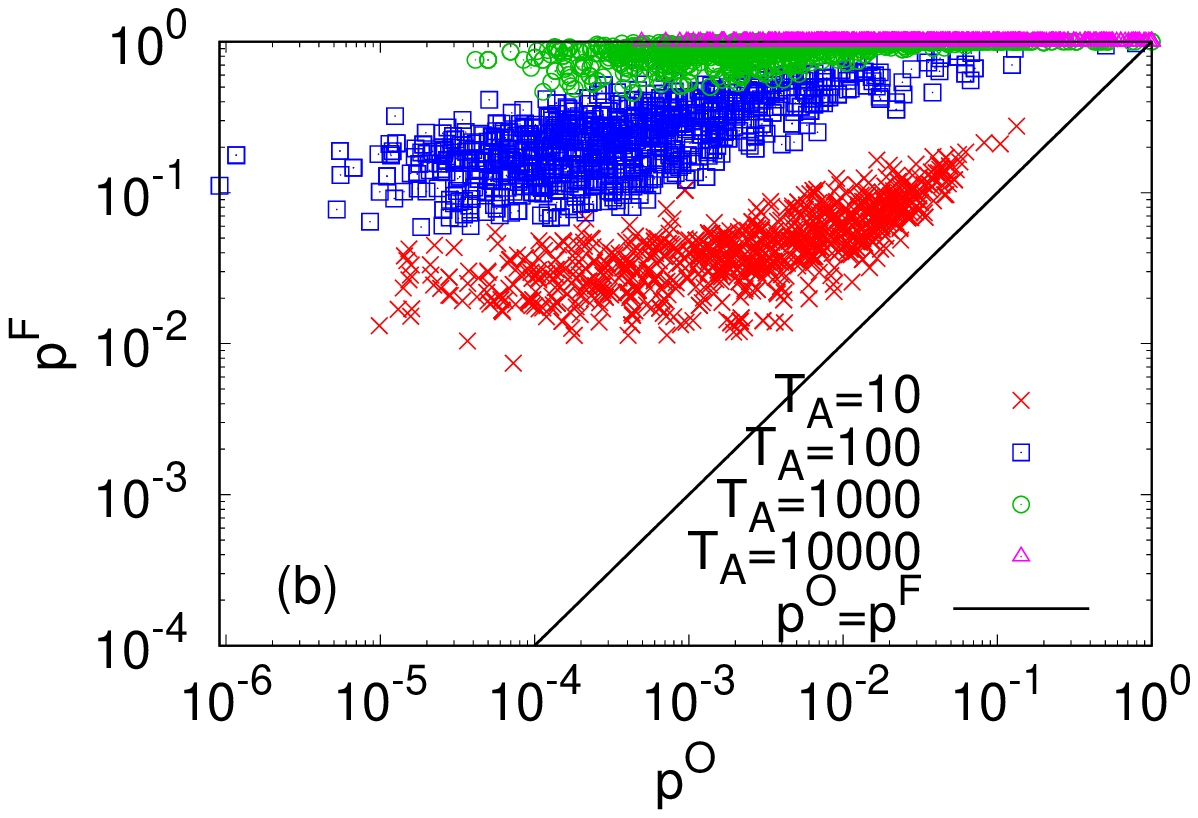}
     \end{minipage}
     \begin{minipage}[r]{0.3\textwidth}
         \includegraphics[scale=0.47]{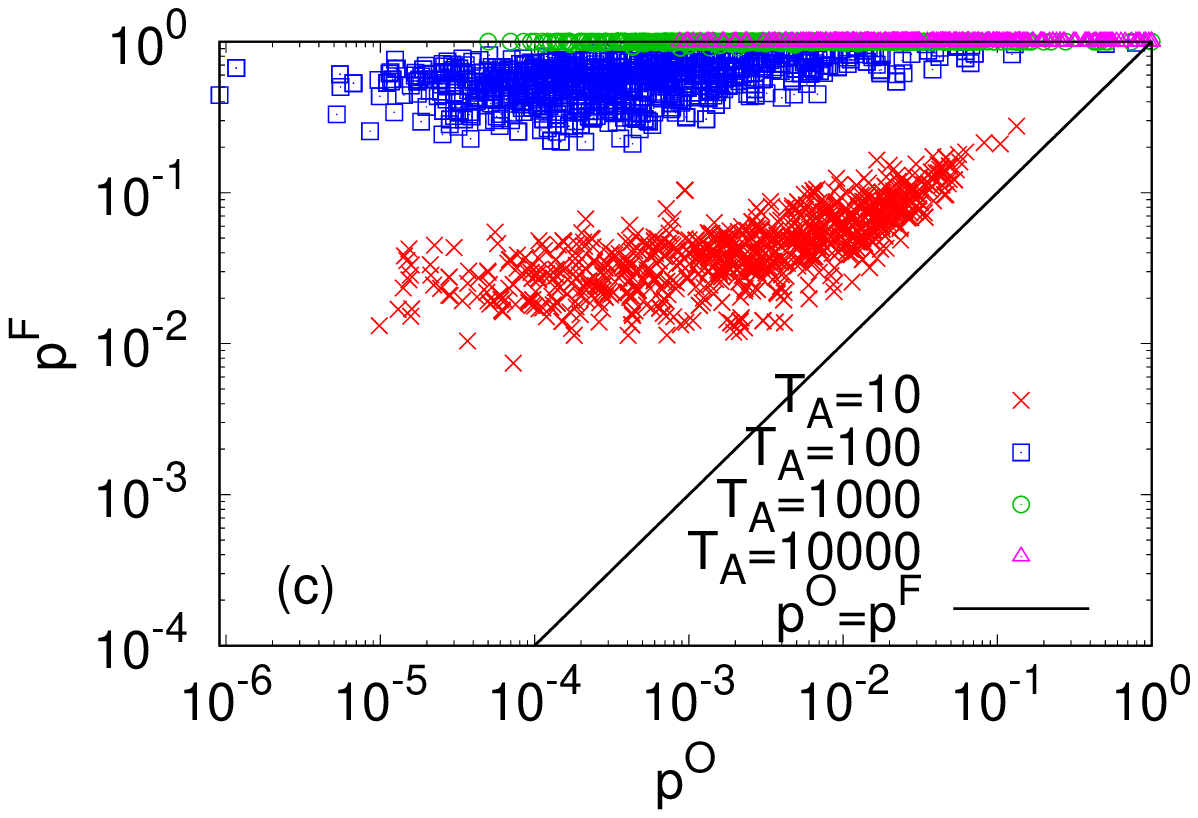}
     \end{minipage}
    \caption{(Color online) Success probability $p^F$ for the 18-variable 2-SAT problems after adding the ferromagnetic trigger Hamiltonian to the Hamiltonian~(\ref{eq_annealing}) versus success probability $p^O$ without adding the trigger Hamiltonian and this for trigger strengths (a) g=0.5, (b) g=1.0 and (c) g =2.0.}
    \label{fig:ferro_success}
\end{figure*}

The effect of adding the ferromagnetic trigger is an indisputable increase in the minimum energy gap of the problems as shown in Fig.~\ref{fig:ferro_mingap}. Here, the minimum energy gap with the ferromagnetic trigger has been marked as $\Delta^{F}$. The number of anticrossings between the ground state and the first-excited state is one. This leads to an expected enhancement in the resulting success probability after adding the trigger, for all annealing times as shown in Fig.~\ref{fig:ferro_success}, except for 17 problems with $T_A=10$. For convenience, the success probability for the Hamiltonian after adding the ferromagnetic trigger Hamiltonian is labeled $p^F$. The enlargement of the minimum energy gap and the improvement in the success probability increases with increasing strength of the ferromagnetic trigger. It is also noted that adding the trigger shifts the occurrence of the minimum energy gap to larger $s$ values. 

\begin{figure*}[ht]
    \centering
          \begin{minipage}[l]{0.32\textwidth}
         \includegraphics[scale=0.47]{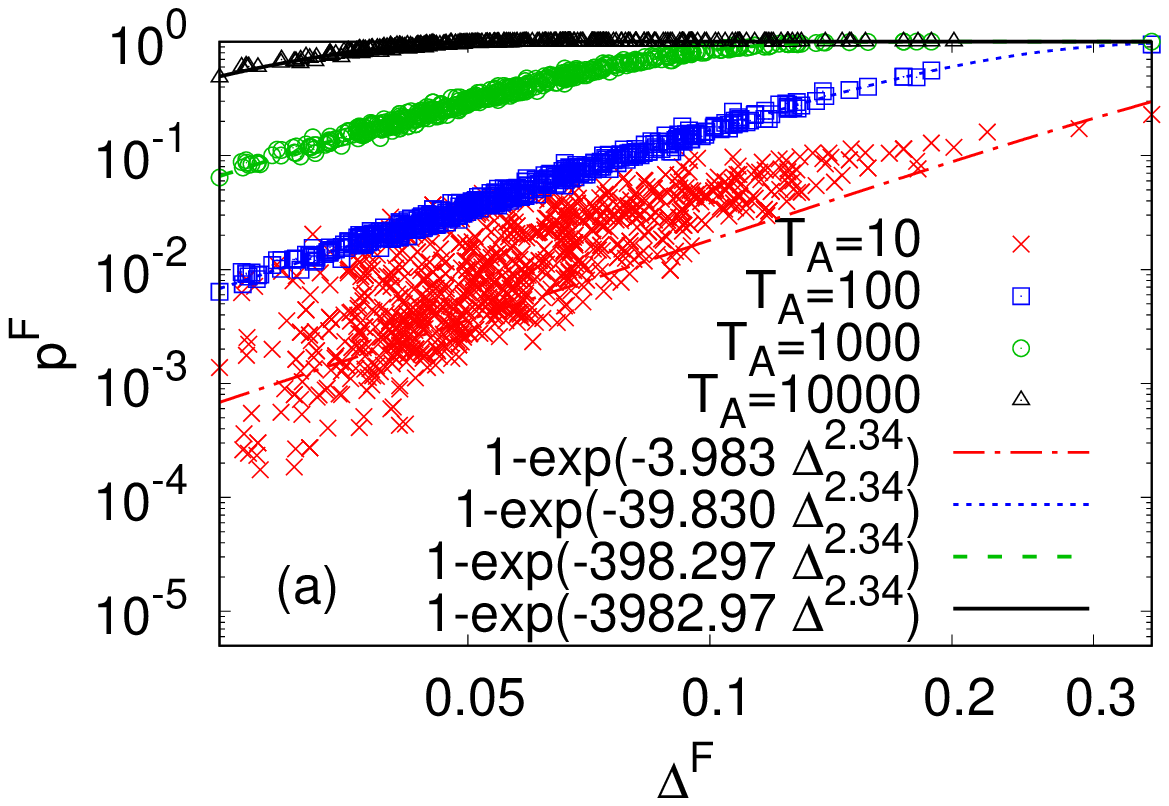}
     \end{minipage}
     \begin{minipage}[c]{0.32\textwidth}
         \includegraphics[scale=0.47]{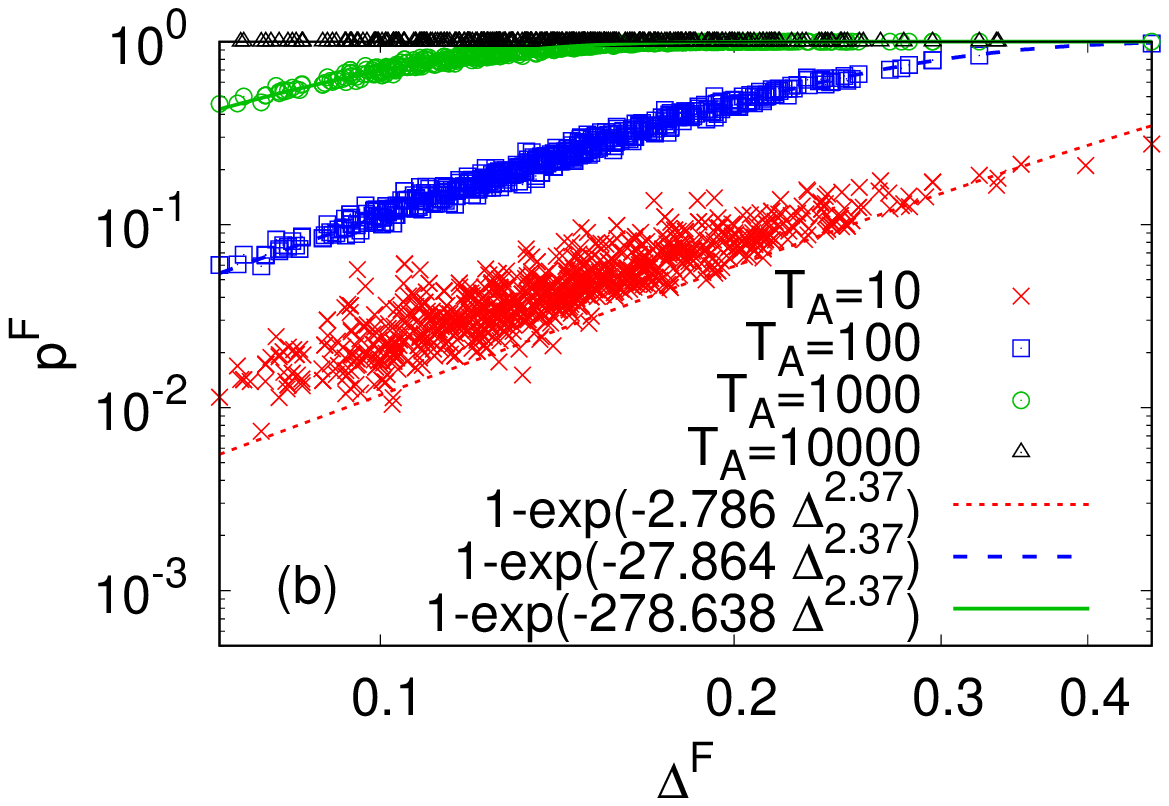}
     \end{minipage}
     \begin{minipage}[r]{0.32\textwidth}
         \includegraphics[scale=0.47]{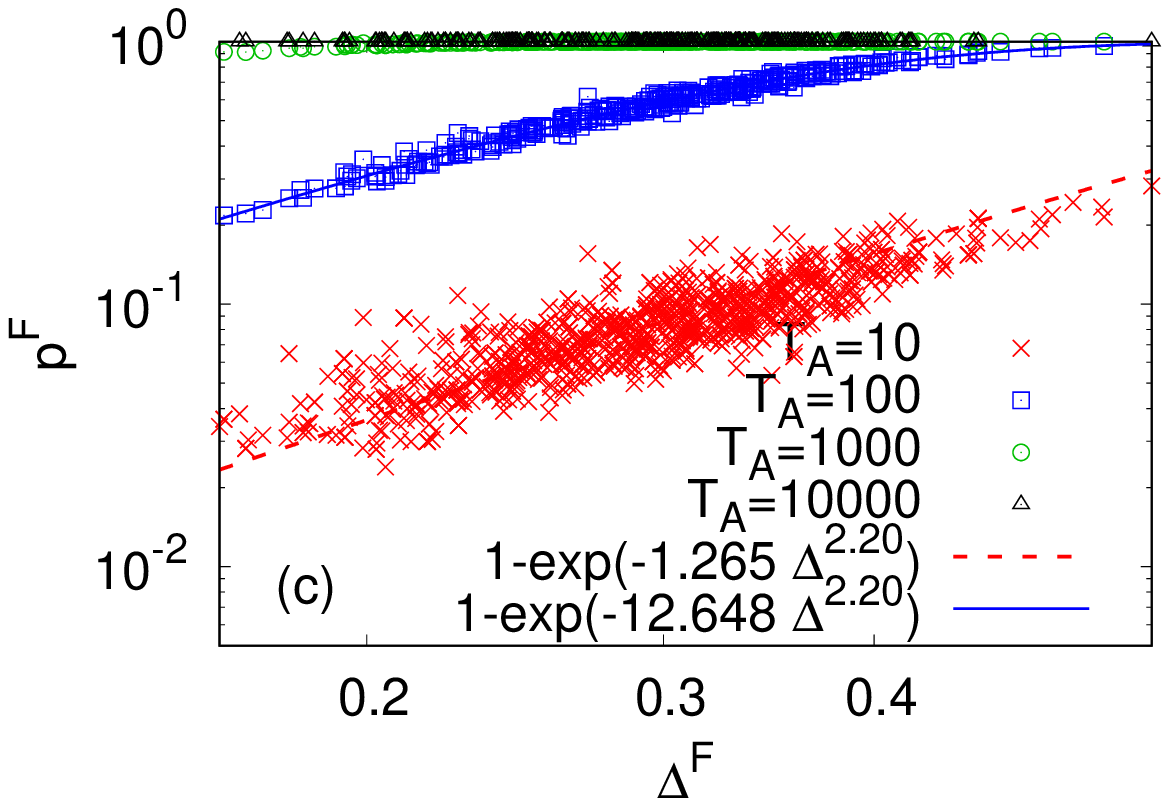}
     \end{minipage}
    \caption{(Color online) Success probability $p^F$ versus minimum energy gap $\Delta^F$ for 18-variable 2-SAT problems after adding ferromagnetic trigger Hamiltonian to the Hamiltonian~(\ref{eq_annealing}) with trigger strength (a) g=0.5, (b) g=1.0 and (c) g=2.0 and for various annealing times $T_A$. The lines are fits to the data.}
    \label{fig:Ferro_succvsgap}
\end{figure*}

Figure \ref{fig:Ferro_succvsgap}  shows the success probability $p^F$ as a function of the minimum energy gap $\Delta^F$ for various annealing times $T_A$ for the 18-variable 2-SAT problems, after having added the ferromagnetic trigger Hamiltonian to the Hamiltonian~(\ref{eq_annealing}). As in the previous section, we have fit the data by using the function $p=1-\exp({-a\Delta^{b}})$ for the data obtained for $T_A=10000$ for $g=0.5$, and $T_A=1000$ for $g=1.0,2.0$. The parameter $a$ is successively scaled down by a factor of ten when the annealing time is scaled down by a factor of ten. The parameter $b$ is found to lie in the range of 2.20 to 2.37. The scattering of the data points is relatively large for $T_A=10$, hinting at deviations from the Landau-Zener model due to a short annealing time. The scattering becomes insignificant for larger annealing times, and the underlying evolution of the state under the action of the ferromagnetic trigger can be described by the Landau-Zener model. Since $\Delta^F$ becomes increasingly large as the trigger strength is increased, the spread of the corresponding points decreases successively, such that for $g=2.0$, and $T_A=1000,10000$ almost all the problems have a success probability close to 1. \\

\subsection{With the antiferromagnetic trigger Hamiltonian}
\label{sec:antiferro}

\begin{figure*}[ht]
    \centering
    \begin{minipage}[l]{0.32\textwidth}
         \includegraphics[scale=0.47]{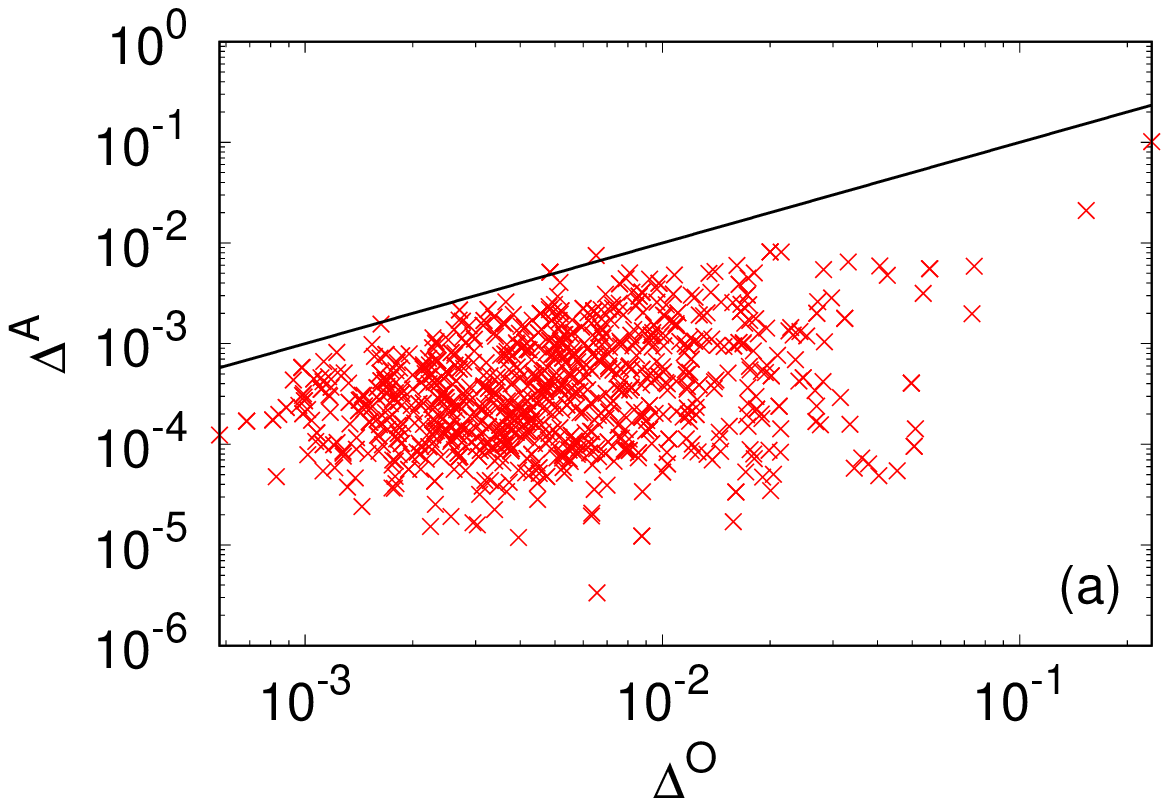}
     \end{minipage}
     \begin{minipage}[c]{0.32\textwidth}
         \includegraphics[scale=0.47]{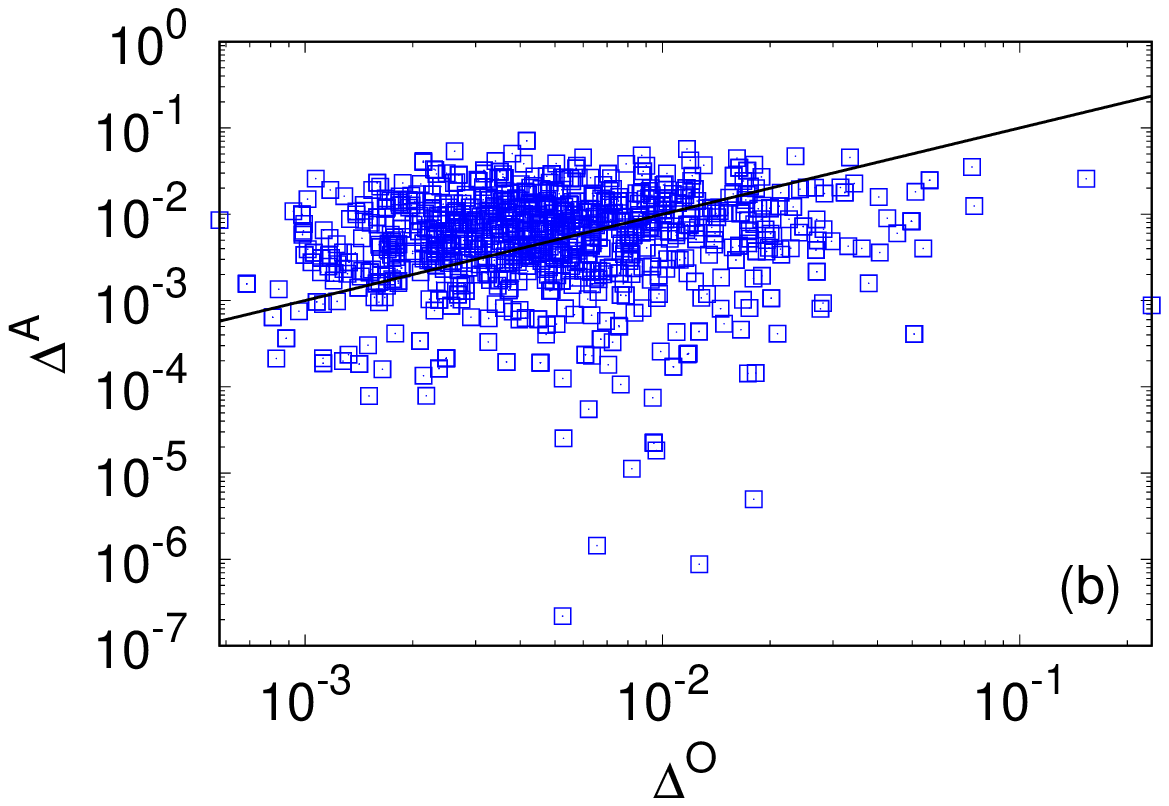}
     \end{minipage}
     \begin{minipage}[r]{0.3\textwidth}
         \includegraphics[scale=0.47]{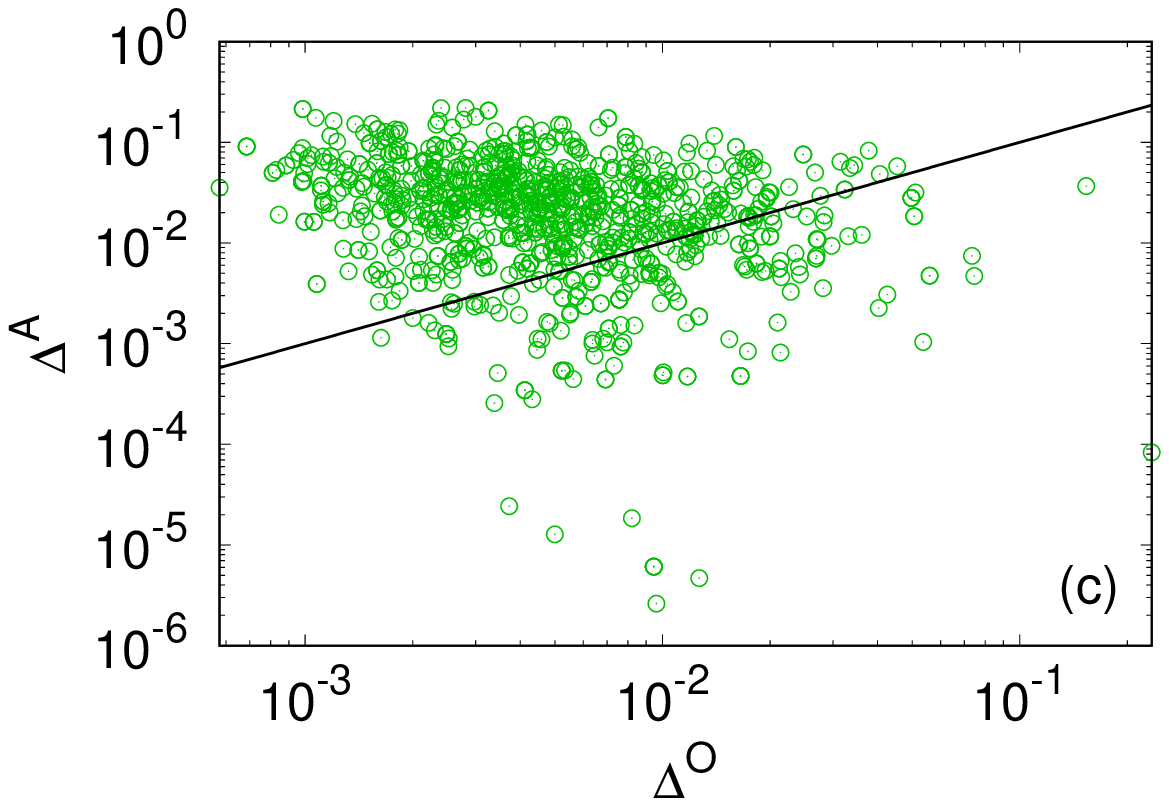}
     \end{minipage}
    \caption{(Color online) Minimum energy gaps $\Delta^A$ of the 18-variable 2-SAT problems after adding the antiferromagnetic trigger Hamiltonian to the Hamiltonian~(\ref{eq_annealing}) versus minimum energy gap $\Delta^O$ without adding the trigger Hamiltonian and this for trigger strengths (a) $g$=0.5, (b) $g$=1.0 and (c) $g$ =2.0.}
    \label{fig:antiferro_gaps}
\end{figure*}
Adding the antiferromagnetic trigger Hamiltonian to the Hamiltonian~(\ref{eq_annealing}) leads to quite distinct and interesting results for the performance of quantum annealing. The minimum energy gaps after adding the antiferromagnetic trigger are labeled $\Delta^A$, while the resulting success probabilities are referred to as $p^A$.\\

The overall result of adding the antiferromagnetic trigger is a reduction in the minimum energy gap for a majority of the 12-variable problems. As the strength of the trigger increases, the number of problems with an enlarged minimum energy gap increases (see Table~\ref{tab:antiferro_largegap}). In comparison, there are fewer cases with reduced minimum energy gaps for the 18-variable problems, especially for $g=1.0$ and $g=2.0$. Figure~\ref{fig:antiferro_gaps} shows the scatter plots of the minimum energy gaps $\Delta^A$ after adding the antiferromagnetic trigger versus minimum energy gaps $\Delta^O$ without adding the trigger Hamiltonian.
\begin{table}
\begin{center}
\caption{Number of 12-variable and 18-variable 2-SAT problems with enlarged minimum energy gaps after adding the antiferromagnetic trigger with strengths $g$ to the Hamiltonian~(\ref{eq_annealing}). Both sets consist of 1000 problems each.}
\begin{tabular}{ |c|c|c| } 
 \hline
 \bm{$g$} & \textbf{12-variables} & \textbf{18-variables} \\ 
 \hline
 0.5 & 1 & 4 \\ 
 \hline
 1.0 & 120 & 565 \\ 
 \hline
 2.0 & 202 & 792 \\ 
 \hline
\end{tabular}
\label{tab:antiferro_largegap}
\end{center}
\end{table}

Another result of adding the antiferromagnetic trigger is a distortion of the energy spectra of the problems in ways such that the number of anticrossings between the ground state and the first-excited state of the Hamiltonian increases with increasing $g$ (see Table~\ref{tab:anticrossings}). Moreover, for some problems, it is observed that the anticrossing between these states gets stretched (i.e., it extends for a longer range of the annealing parameter $s$\textcolor{CV}{, see, e.g., Fig.~\ref{fig:stretched}(a)}). These changes may give rise to several nonadiabatic transitions that may increase the chances of success for these problems. Therefore, in contrast with the ferromagnetic trigger, the antiferromagnetic trigger induces changes that vary greatly with the trigger strength, the annealing time and the specific problem. Furthermore, the addition of the antiferromagnetic trigger shifts the occurrence of the minimum energy gaps to smaller $s$ values. \\

\begin{table}
\begin{center}
\caption{Number of 12-variable and 18-variable 2-SAT problems with different numbers of anticrossings, $N_A$, after adding the antiferromagnetic trigger Hamiltonian with strengths $g$, to the Hamiltonian~(\ref{eq_annealing}). Both sets consist of 1000 problems each.}
\resizebox{\columnwidth}{!}{
\begin{tabular}{ |c||c|c|c||c|c|c| } 
 \hline
 &\multicolumn{3}{c||}{12-variables} & \multicolumn{3}{c|}{18-variables}\\
 \hline
 \bm{$N_A$} & \bm{$g=0.5$} & \bm{$g=1.0$} & \bm{$g=2.0$}  & \bm{$g=0.5$} & \bm{$g=1.0$} & \bm{$g=2.0$} \\ 
 \hline
 1 & 923 & 202 & 1 & 796 & 37 & 0 \\ 
 \hline
 2 & 77 & 705 & 132 & 204 & 394 & 24 \\ 
 \hline
 3 & 0 & 92 & 439 & 0 & 468 & 182 \\ 
 \hline
 4 & 0 & 1 & 33 & 0 & 99 & 424 \\ 
 \hline
 5 & 0 & 0 & 65 & 0 & 2 & 290 \\ 
 \hline
 6 & 0 & 0 & 0 & 0 & 0 & 68 \\ 
 \hline
 7 & 0 & 0 & 0 & 0 & 0 & 12 \\ 
 \hline
\end{tabular}
}
\label{tab:anticrossings}
\end{center}
\end{table}

\begin{figure*}
    \centering
     \begin{minipage}[l]{0.32\textwidth}
         \includegraphics[scale=0.47]{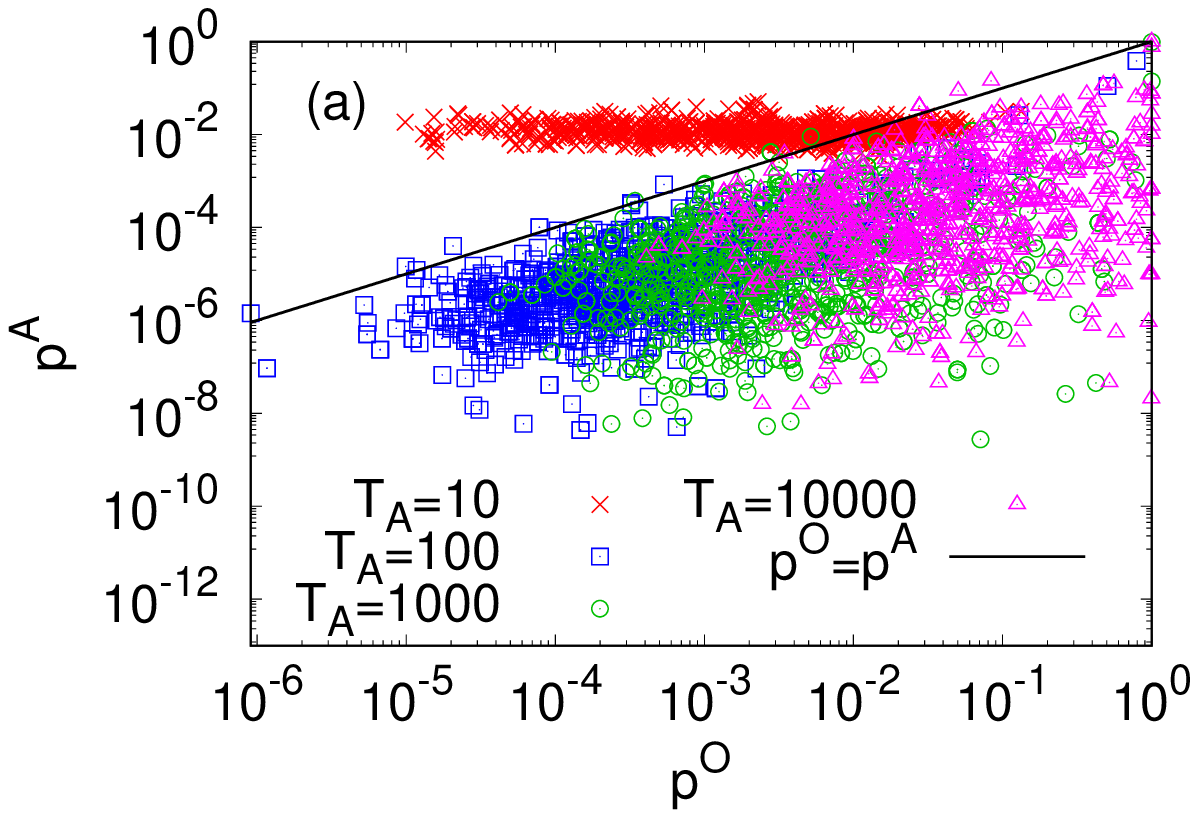}
     \end{minipage}
     \begin{minipage}[c]{0.32\textwidth}
         \includegraphics[scale=0.47]{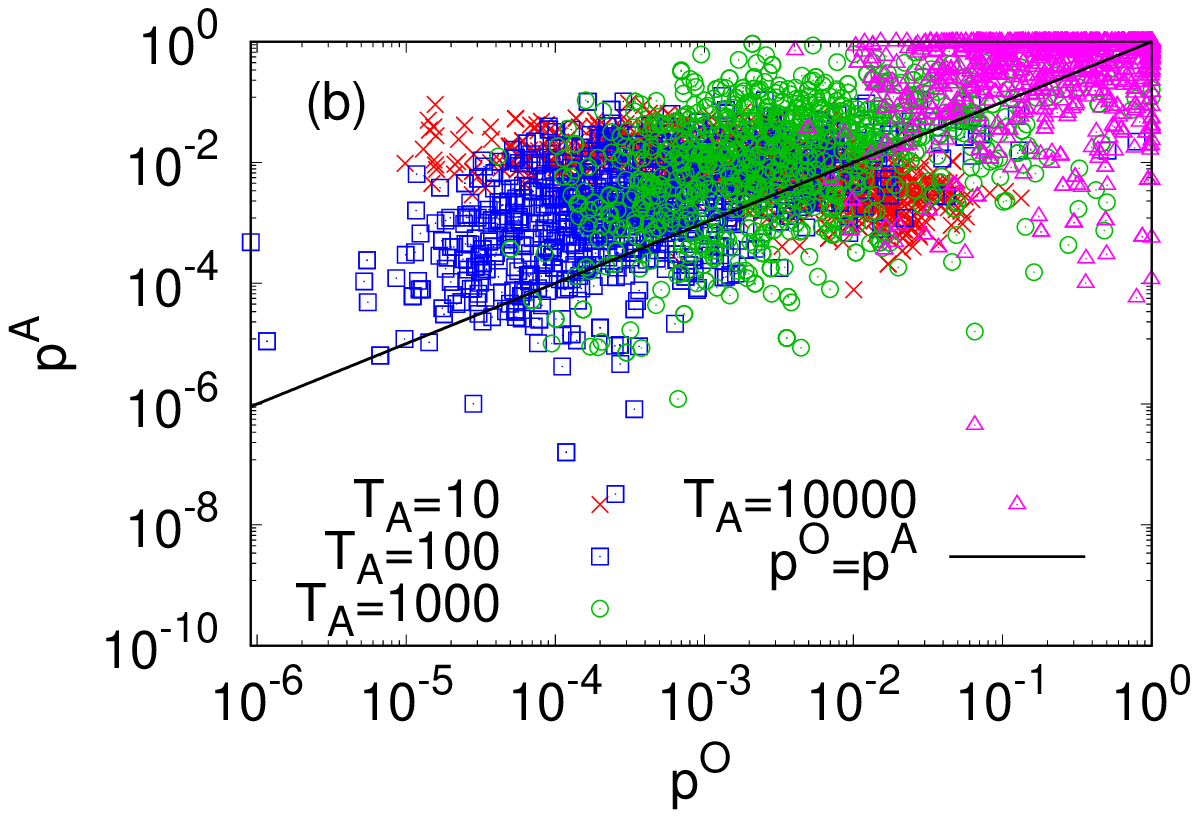}
     \end{minipage}
     \begin{minipage}[r]{0.3\textwidth}
         \includegraphics[scale=0.47]{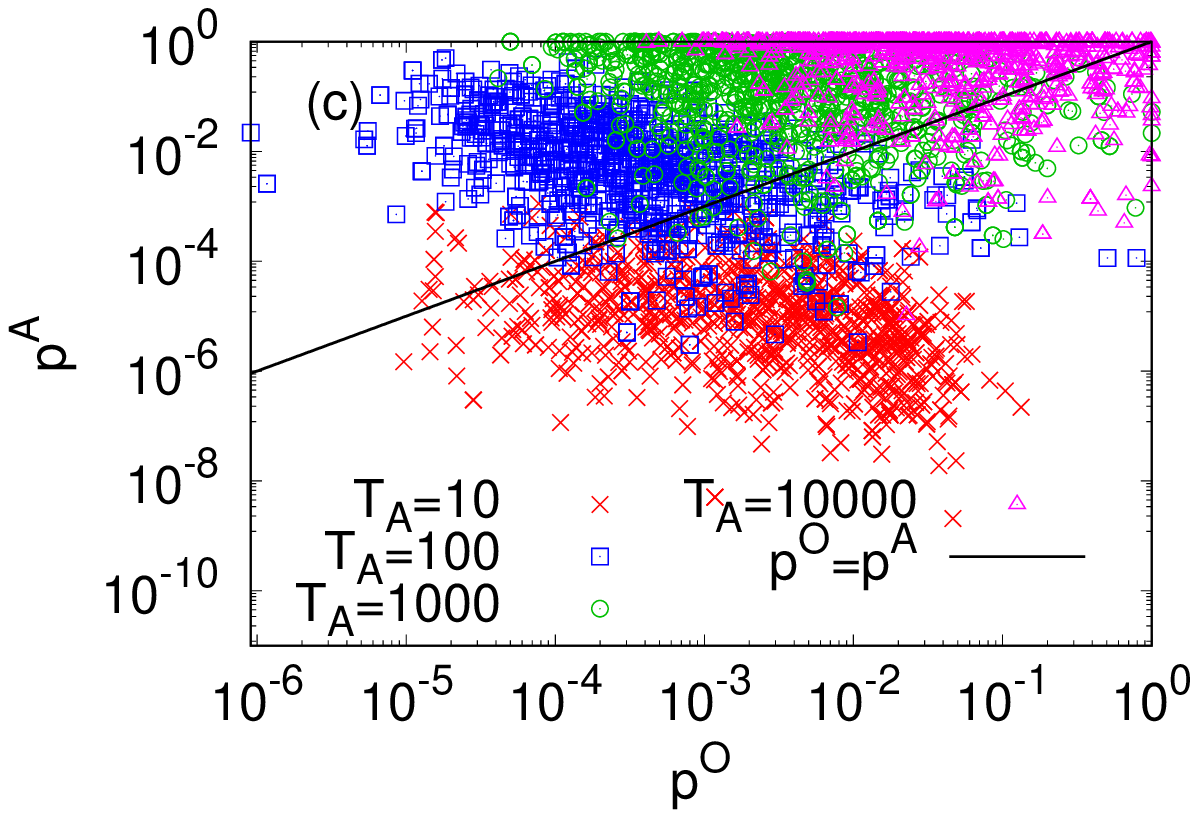}
     \end{minipage}
    \caption{(Color online) Success probability $p^A$ of the 18-variable 2-SAT problems after adding the antiferromagnetic trigger Hamiltonian to the Hamiltonian~(\ref{eq_annealing}) versus success probability $p^O$ without adding the trigger Hamiltonian and this for trigger strengths (a) $g$=0.5, (b) $g$=1.0 and (c) $g$ =2.0 and various annealing times $T_A$.}
     \label{fig:antiFerro_success}
\end{figure*}

Table~\ref{tab:antiferro_largesuccess} shows the number of 12-variable and 18-variable 2-SAT problems with a larger success probability after adding the antiferromagnetic trigger to the Hamiltonian~(\ref{eq_annealing}). The scatter plots for the success probability for the 18-variable problems are shown in Fig.~\ref{fig:antiFerro_success}. It is seen that for $g=0.5$, a short annealing time of $T_A=10$ greatly improves the success probability for the difficult cases. As the trigger strength $g$ increases, the resulting enhancements are mainly for longer annealing times.

\begin{table}
\begin{center}
\caption{ Number of 12-variable and 18-variable 2-SAT problems with increased success probabilities after adding the antiferromagnetic trigger with strengths $g$, to the Hamiltonian~(\ref{eq_annealing}). Both sets consist of 1000 problems each.}
\resizebox{\columnwidth}{!}{
\begin{tabular}{ |c||c|c|c||c|c|c| } 
 \hline
 &\multicolumn{3}{c||}{12-variables} & \multicolumn{3}{c|}{18-variables}\\
 \hline
 \textbf{$g$} & \bm{$T_A=10$} & \bm{$T_A=100$} & \bm{$T_A=1000$}  & \bm{$T_A=10$} & \bm{$T_A=100$} & \bm{$T_A=1000$} \\ 
 \hline
 0.5 & 439 & 2 & 2 & 732 & 10 & 3 \\ 
 \hline
 1.0 & 377 & 215 & 200 & 627 & 798 & 992 \\ 
 \hline
 2.0 & 15 & 158 & 225 & 41 & 754 & 844 \\ 
 \hline
\end{tabular}
}
\label{tab:antiferro_largesuccess}
\end{center}
\end{table}

\begin{figure*}
    \centering
     \begin{minipage}[l]{0.32\textwidth}
         \includegraphics[scale=0.47]{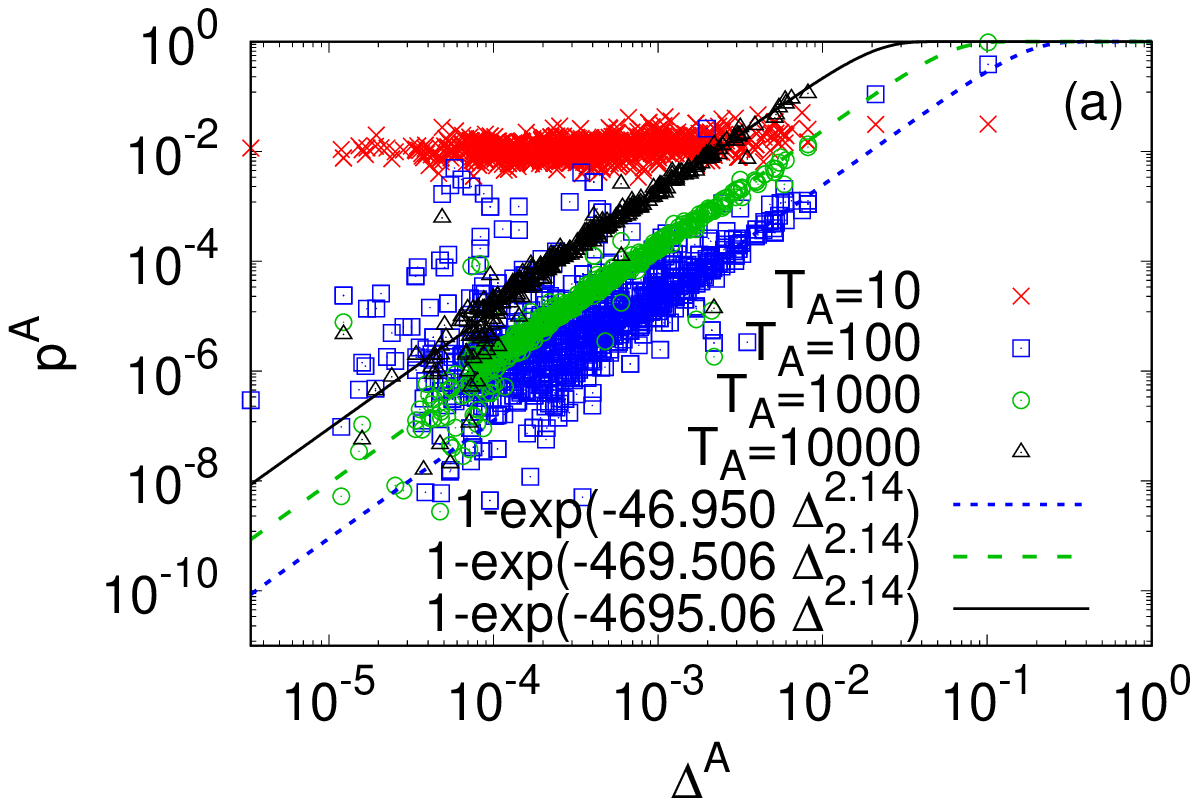}
     \end{minipage}
     \begin{minipage}[c]{0.32\textwidth}
         \includegraphics[scale=0.47]{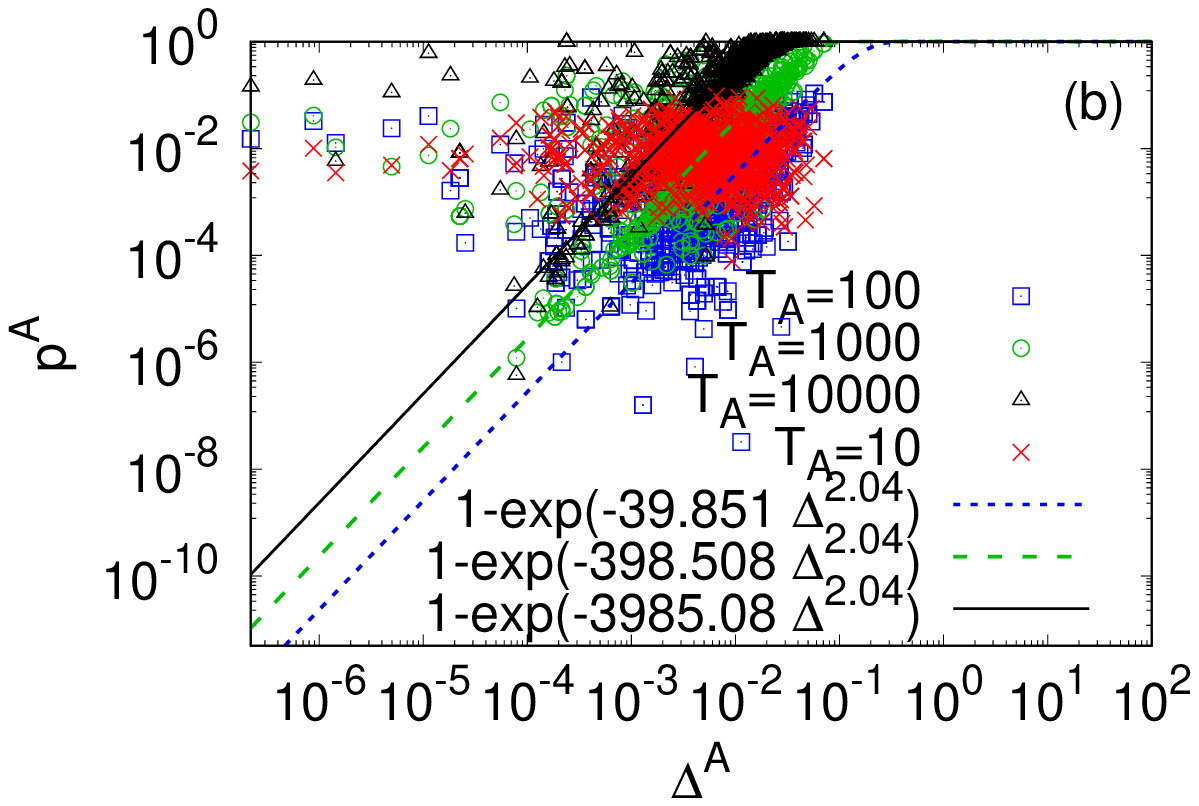}
     \end{minipage}
     \begin{minipage}[r]{0.32\textwidth}
         \includegraphics[scale=0.47]{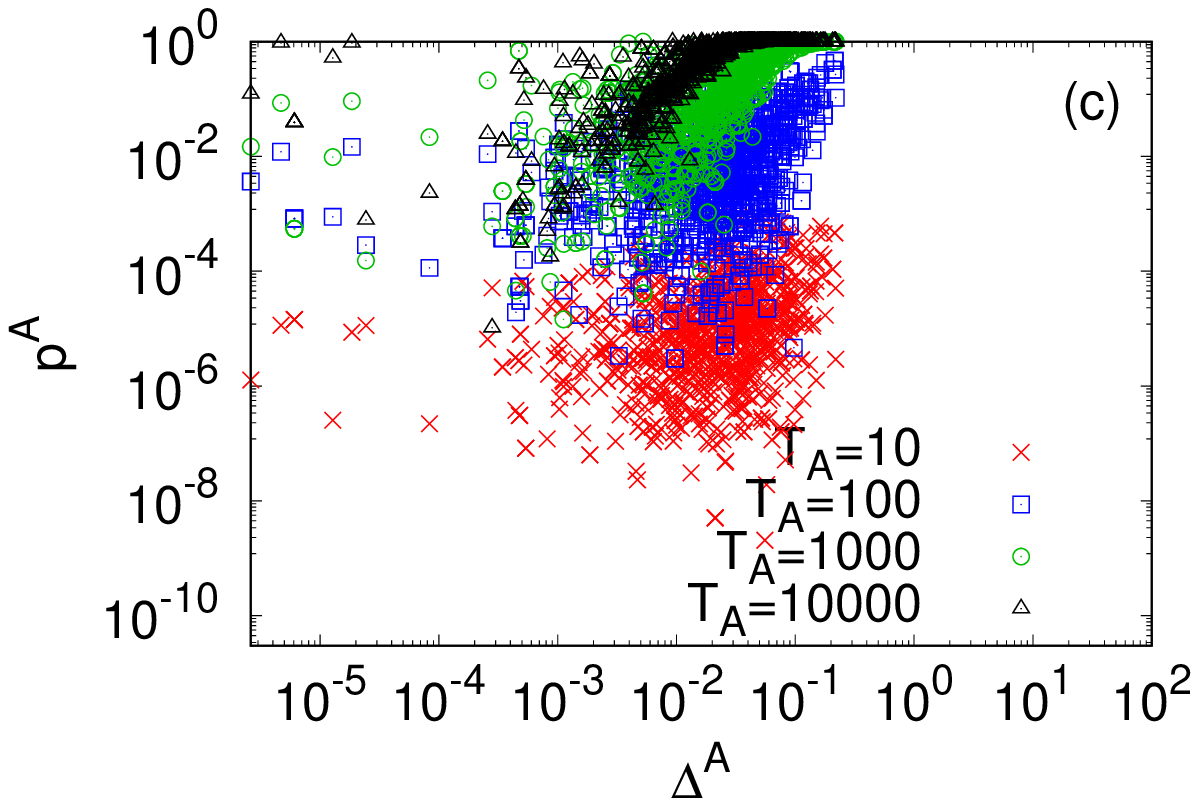}
     \end{minipage}
    \caption{(Color online) Success probability $p^A$ versus minimum energy gap $\Delta^A$ for the 18-variable 2-SAT problems after adding the antiferromagnetic trigger Hamiltonian to the Hamiltonian~(\ref{eq_annealing}) with strength (a) $g$=0.5, (b) $g$=1.0, and (c) $g$=2.0 and for various annealing times $T_A$. The lines are fits to the data.}
     \label{fig:antiFerro_succvsgap}
\end{figure*}

To understand these observations, we focus on the scatter plots shown in Fig.7 that depict the success probability $p^A$ as a function of the minimum energy gap $\Delta^A$ for the 18-variable 2-SAT problems. The fitting curves are obtained by using the same method as described in the previous section. However, for some parameter settings we have omitted a fit to the data because of the large scattering in the data. For $g=0.5$ and $T_A=10$ the resulting fitted curve (not shown) is rather flat, indicating that for these parameters, adding the antiferromagnetic trigger leads to large deviations from the Landau-Zener model. This behavior is similar to the one if noise is acting on a system, e.g., the coupling of the system with a heat bath can cause transitions between the ground state and higher energy states, at points other than the $s$ value corresponding to the anticrossing.

If the annealing time \textcolor{CV}{increases}, the success probabilities follow the Landau-Zener prediction well, as can be observed from Fig.~\ref{fig:antiFerro_succvsgap}. Upon increasing the trigger strength $g$, the flatness of the resulting fitted curve (omitted in the figure) for $T_A=10$ decreases.
This can be explained by the enlargement of the minimum energy gaps for a larger number of problems with increasing $g$ values. However, the scattering of the points around the fitted curves increases with an increase in the trigger strength, which can be attributed to the increase in the number of anticrossings between the ground state and the first-excited state of the Hamiltonian as given in Table~\ref{tab:anticrossings}.

Finally, we discuss the nonadiabatic mechanisms, arising from the distortion of the energy spectrum caused by adding the antiferromagnetic trigger, which can prove to be beneficial for enhancing the success probability for some problems.
\begin{itemize}

\item \textbf{Even number of comparably small anticrossings:} The most beneficial mechanism is the presence of an even number of small and comparable anticrossings or crossings between the ground state and the first-excited state of the Hamiltonian \cite{albash2018adiabatic,somma2012quantum}. For our problem sets, most such cases had just two of these useful anticrossings or crossings, although this combination was rather rare. In such cases, even if the annealing time is too short for the state of the system to follow the ground state, and thus causing it to deviate from the ground state at the first anticrossing, the presence of the second anticrossing increases the chances of the state of the system to return to the ground state. \\
An example of this can be seen in Fig.~\ref{fig:2_anticrossings} for a 12-variable problem numbered as problem ``709" with $g=2.0$ and $T_A=100$. \textcolor{CV}{In Appendix~\ref{sec:appendix_D}, we make available the 2-SAT problem 709, and other individual problem instances discussed in the paper.} As can be observed from Fig.~\ref{fig:2_anticrossings}(a), there are two crossings between the ground state and the first-excited state, namely at $s=0.165$ and $s=0.248$. From the overlap of the state of the system with the three lowest lying eigenstates of the instantaneous Hamiltonian [Fig.~\ref{fig:2_anticrossings}(b)] it can be clearly observed that the amplitude in the ground state is transferred to the first-excited state at $s=0.165$, but most of this amplitude is transferred to the ground state at $s=0.248$, increasing the success probability for this case.

\begin{figure*}[ht]
    \centering
     \begin{minipage}[l]{0.49\textwidth}
         \includegraphics[scale=0.7]{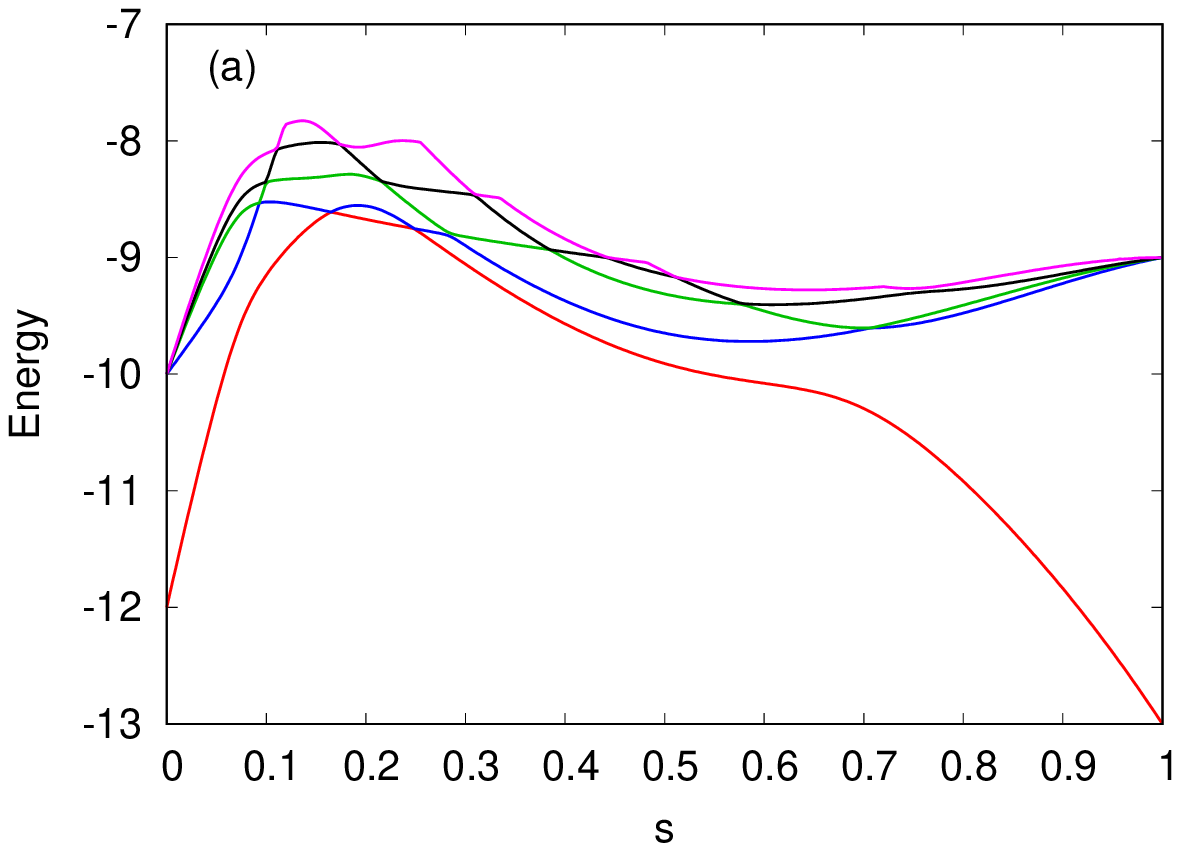}
     \end{minipage}%
     \begin{minipage}[r]{0.49\textwidth}
         \centering
         \includegraphics[scale=0.7]{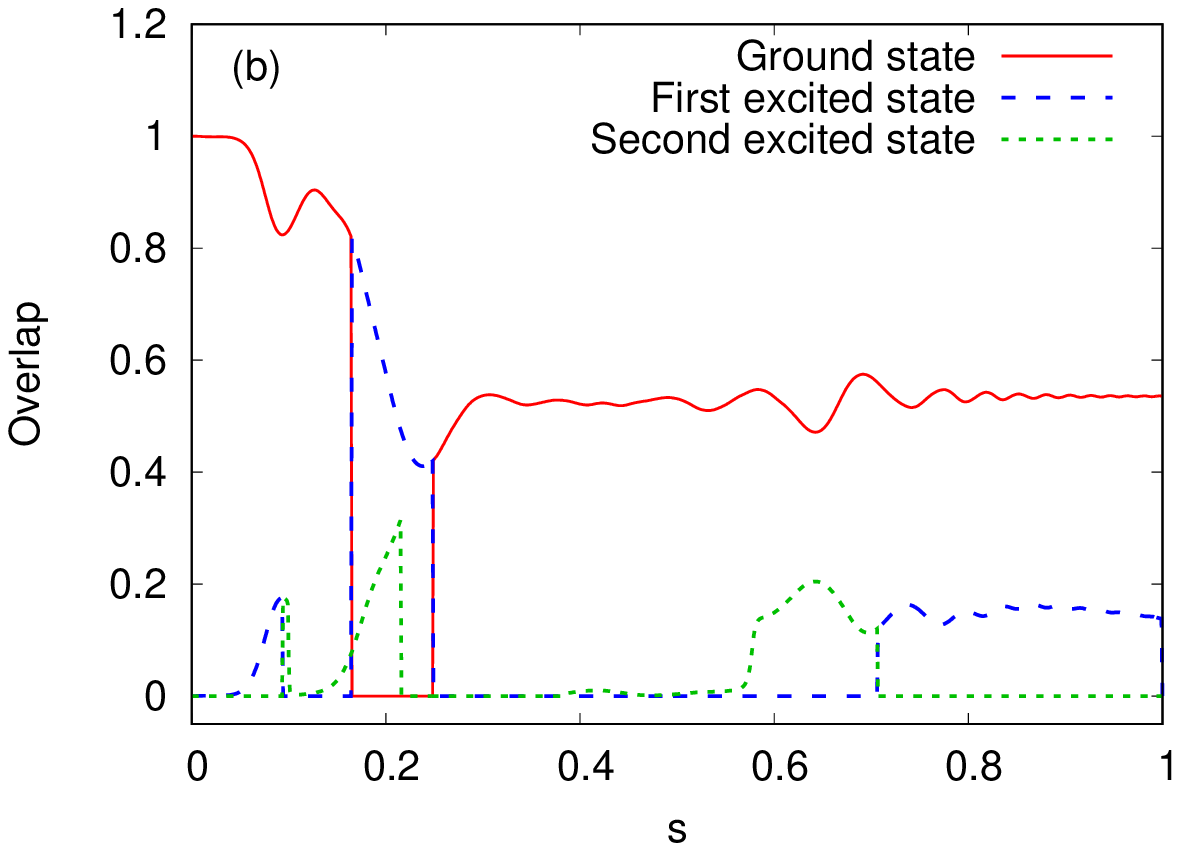}
     \end{minipage}
     \caption{(Color online) (a) Energy spectrum of the lowest lying states represented by solid lines and (b) the overlap of the state of the system with the instantaneous low energy states for $T_A$=100 for the 12-variable 2-SAT problem number 709 and with $g$=2.0.}
     \label{fig:2_anticrossings}
\end{figure*}

    \item \textbf{Fast annealing: }For short annealing times, the state of the system deviates from the ground state even before reaching the anticrossing \cite{albash2018adiabatic,crosson2014different}. Then, upon reaching the anticrossing, there is a finite possibility that part of the amplitude returns to the ground state. Figure~\ref{fig:returning} shows one such 12-variable problem numbered as ``950" for $g=0.5$ and $T_A=10,100$. It can be observed that for $T_A=100$ the state of the system remains in the ground state until it reaches the anticrossing, from where it shifts to the first-excited state. However, for a shorter annealing time, the state of the system already deviates from the ground state before the anticrossing, but comes again close to it at the anticrossing, thereby increasing the overlap of the final state with the ground state.\\
    This mechanism is also observed in the quantum approximate optimization algorithm (QAOA) with intermediate $p$ levels for MaxCut problems, if the optimal parameters are interpreted as the scheme for quantum annealing \cite{zhou2020quantum}. In general, using the scheme for quantum annealing as the initialization parameters for the QAOA can be advantageous for its performance \cite{willsch2020benchmarking,sack2021quantum}, and similarly, using the optimal parameters from the QAOA as the annealing scheme for quantum annealing can enhance the success probability for short annealing times.
    
\begin{figure}[ht]
    \centering
    \includegraphics[scale=0.7]{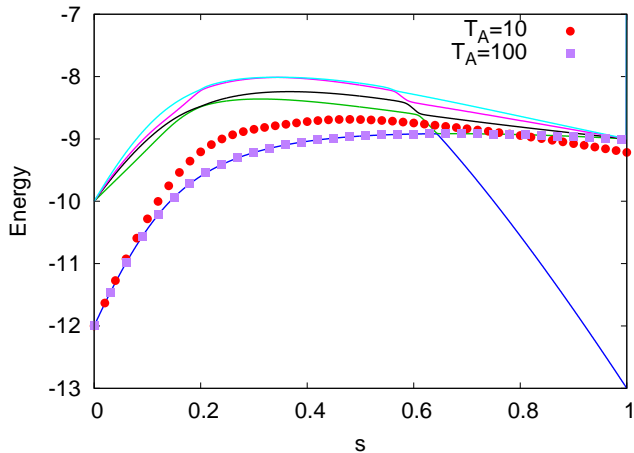}
    \caption{(Color online) Energy spectrum of the lowest lying states for the 12-variable problem number 950 with $g=0.5$. The solid lines represent the energy spectrum of the lowest lying states of the problem while the points represent the average energy of the instantaneous state of the system for $T_A=10,100$. }
    \label{fig:returning}
\end{figure}
    
    \item \textbf{Stretching of the anticrossing: }The energy spectrum changes in such a way that the ground state and the first-excited state stay in close vicinity for a longer range of the annealing parameter $s$. According to Eq.~(\ref{eq:adiabatic}), the minimum time required to ensure that the evolution remains adiabatic depends on two factors: the minimum energy gap $\Delta$, and the rate of change of the Hamiltonian with respect to the annealing parameter. Upon adding the antiferromagnetic trigger to the Hamiltonian~(\ref{eq_annealing}),, the shape of the anticrossing becomes distinct from the rest of the problems in the set \textcolor{CV}{(which have sharp anticrossings limited to a small range of $s$)}, thus affecting the success probability for the chosen annealing time.
    Therefore, despite of a smaller minimum energy gap, in some cases, the success probability improves due to the stretching of the anticrossing. Additionally, this stretching can cause the amplitude of the wavefunction to oscillate between these states, which might, coincidentally, result in a larger overlap of the final state with the ground state at the end of the annealing process. An example is shown in Fig.~\ref{fig:stretched} for the problem with 12 variables numbered as ``103" with $g=2.0$ and $T_A=100$. In this case, the oscillations of the amplitude of the state of the Hamiltonian result in a larger success probability.
\begin{figure*}[ht]
    \centering
     \begin{minipage}[l]{0.49\textwidth}
         \includegraphics[scale=0.7]{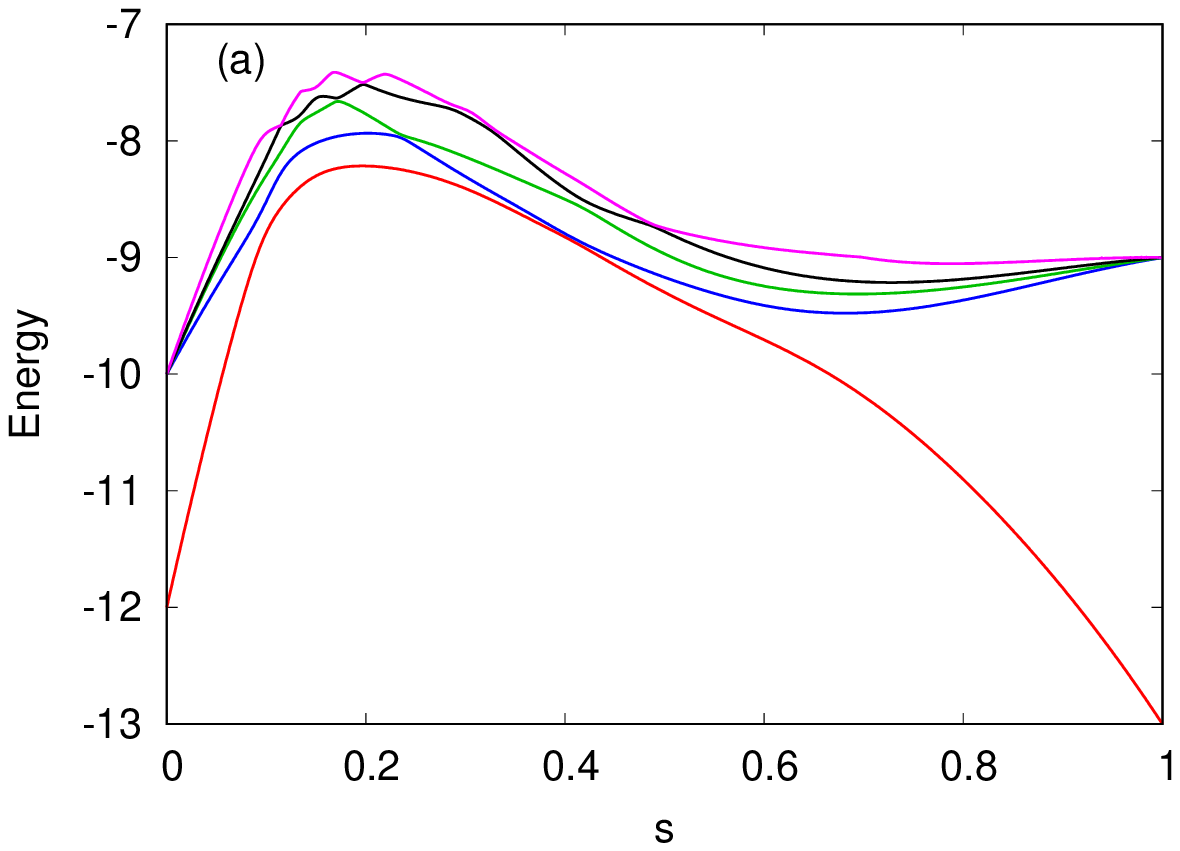}
     \end{minipage}%
     \begin{minipage}[r]{0.49\textwidth}
         \centering
         \includegraphics[scale=0.7]{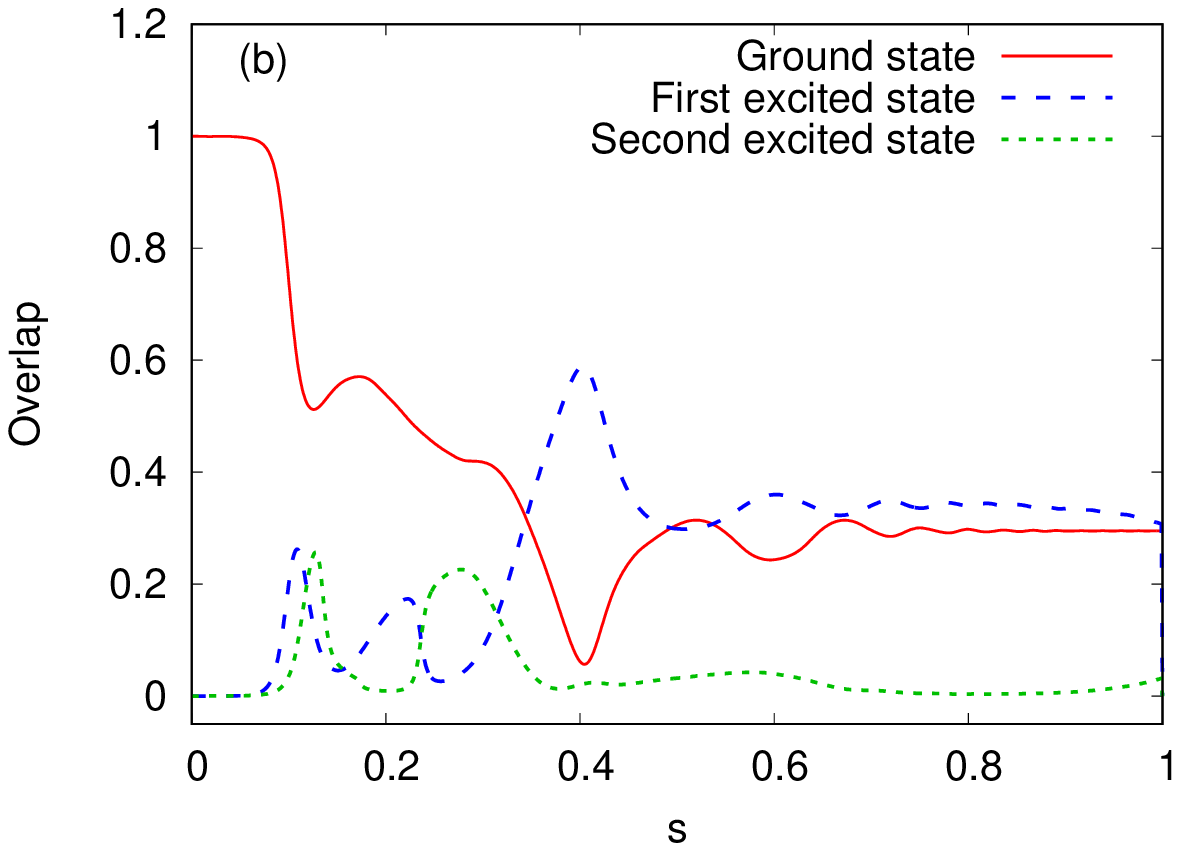}
     \end{minipage}
     \caption{(Color online) (a) Energy spectrum of the lowest lying states represented by solid lines and (b) the overlap of the state of the system with the instantaneous low energy states for $T_A$=100 for the 12-variable 2-SAT problem number 103 and with $g$=2.0.}
     \label{fig:stretched}
\end{figure*}
\end{itemize}

\section{Quantum Annealing for Nonstoquastic Problems}
\label{results_nonstoq}

In this section, we address the results obtained for quantum annealing with the nonstoquastic problem Hamiltonian $H_P$. We perform a similar analysis as we did for the 2-SAT problems by adding the ferromagnetic and antiferromagnetic trigger Hamiltonian to the Hamiltonian~(\ref{eq_annealing}). We used the Lanczos algorithm for determining the ground state of these nonstoquastic problems. Here we show only the results for the 18-variable problems. The results for the 12-variable problems are given in Appendix~\ref{sec:appendix_C}. 

\begin{figure}
    \centering
    \includegraphics[scale=0.7]{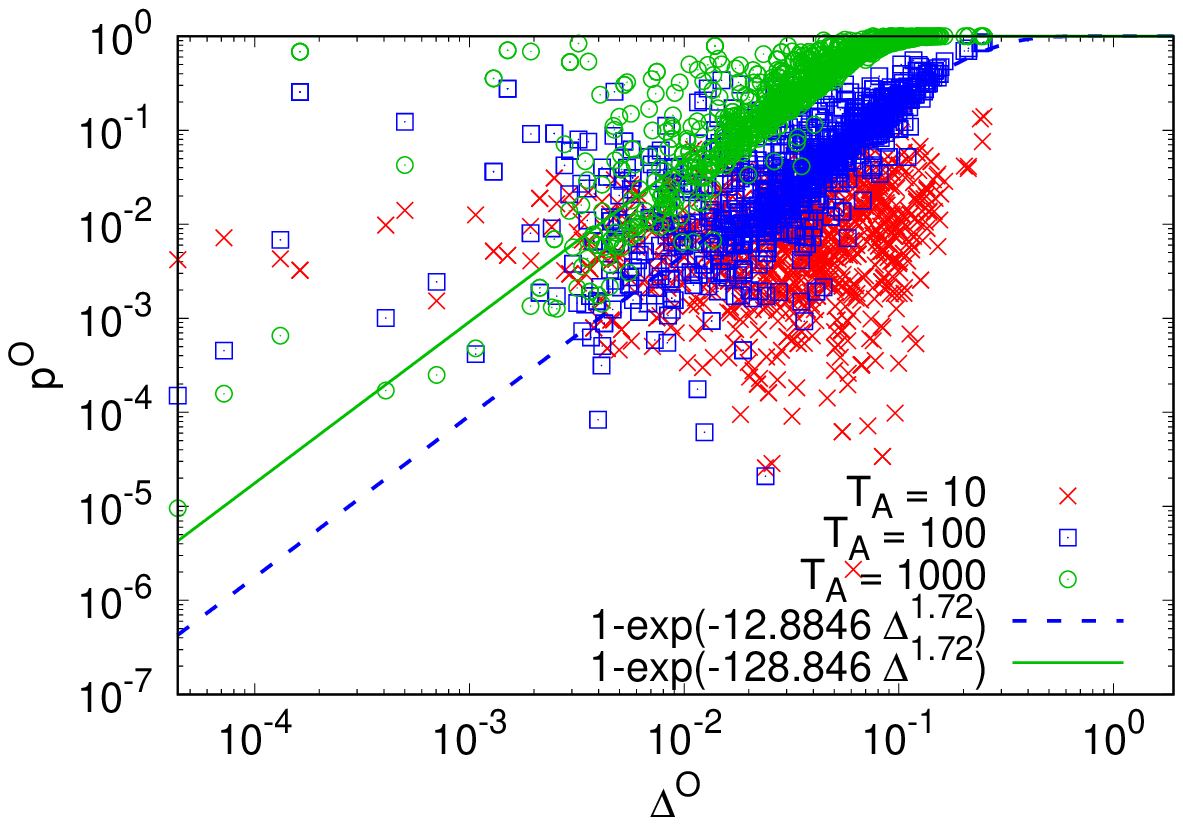}
    \caption{(Color online) Success probability $p^o$ versus minimum energy  gap $\Delta^O$ for 18-variable nonstoquastic problems without adding the trigger Hamiltonian to the Hamiltonian~(\ref{eq_annealing}).}
    \label{fig:Nonstoq_Orig}
\end{figure}

\begin{table}
\begin{center}
\caption{Number of 12-variable and 18-variable nonstoquastic problems with different numbers of anticrossings, $N_A$, without adding the trigger Hamiltonian to the Hamiltonian~(\ref{eq_annealing}). Both sets consist of 1000 problems each.}
\begin{tabular}{ |c|c|c| } 
 \hline
 \bm{$N_A$} & \textbf{12-variables} & \textbf{18-variables} \\ 
 \hline
 1 & 972 & 628 \\ 
 \hline
 2 & 28 & 353 \\ 
 \hline
 3 & 0 & 19 \\ 
 \hline
\end{tabular}
\label{tab:nonstoq_orig_antic}
\end{center}
\end{table}
Figure~\ref{fig:Nonstoq_Orig} shows the success probability $p^O$ versus minimum energy gap $\Delta^O$ for the quantum annealing algorithm without adding the trigger Hamiltonian to the Hamiltonian~(\ref{eq_annealing}). For fitting the data to $p = 1-\exp(-a\Delta^b)$, we used the data points for $T_A=1000$, and we scaled the resulting parameter $a$ as explained in the previous section. Parameter $b$ was found to be 1.72. It can be seen that the data for $T_A=10$ show scattering to such an extent that the tail, corresponding to the smallest energy gaps, appears flat. This indicates that these problems show large deviations from the Landau-Zener model as the annealing time $T_A=10$ is too short. Thus, we conclude that for $T_A=10$ nonadiabatic mechanisms, as in the case of the 2-SAT problems with antiferromagnetic trigger for short annealing times, are dominant for these problems. Table~\ref{tab:nonstoq_orig_antic} shows the number of problems with different number of anticrossings.

We then add the ferromagnetic trigger Hamiltonian to the Hamiltonian~(\ref{eq_annealing}) for solving the nonstoquastic problems. Unlike the set of 2-SAT problems, these problems do not exhibit an enhanced performance upon adding the trigger for all the cases, although, as given in Table~\ref{tab:nonstoq_ferro_mingaps}, a majority of the problems have a larger minimum energy gap after adding the ferromagnetic trigger. This percentage increases with increasing trigger strength. Following a similar trend, the success probability is also improved for a majority of the cases, and the number of problems with enhanced success probability increases with increasing the trigger strength (see Table~\ref{tab:nonstoquastic_ferro_succ}). However, comparing across different annealing times for a given $g$, it can be noted that, in contrast with the implications of the adiabatic theorem, this number decreases as the annealing time is increased. Additionally, it also decreases upon increasing the problem size.

\begin{table}
\begin{center}
\caption{Number of 12-variable and 18-variable nonstoquastic problems with enlarged minimum energy gaps after adding the ferromagnetic trigger with different trigger strengths $g$ to the Hamiltonian~(\ref{eq_annealing}). Both sets consist of 1000 problems each.}
\begin{tabular}{ |c|c|c| } 
 \hline
 \bm{$g$} & \textbf{12-variables} & \textbf{18-variables} \\ 
 \hline
 0.5 & 830 & 656 \\ 
 \hline
 1.0 & 913 & 713 \\ 
 \hline
 2.0 & 986 & 837 \\ 
 \hline
\end{tabular}
\label{tab:nonstoq_ferro_mingaps}
\end{center}
\end{table}

\begin{table}
\begin{center}
\caption{Number of 12-variable and 18-variable nonstoquastic problems with increased success probabilities after adding the ferromagnetic trigger Hamiltonian with strengths $g$ to the Hamiltonian~(\ref{eq_annealing}). Both sets consist of 1000 problems each.}
\resizebox{\columnwidth}{!}{
\begin{tabular}{ |c||c|c|c||c|c|c| } 
 \hline
 &\multicolumn{3}{c||}{12-variables} & \multicolumn{3}{c|}{18-variables}\\
 \hline
 \textbf{$g$} & \bm{$T_A=10$} & \bm{$T_A=100$} & \bm{$T_A=1000$}  & \bm{$T_A=10$} & \bm{$T_A=100$} & \bm{$T_A=1000$} \\ 
 \hline
 0.5  & 859 & 815 & 738 & 722 & 596 & 592 \\ 
 \hline
 1.0 & 925 & 894 & 787 & 851 & 678 & 647 \\ 
 \hline
 2.0 & 961 & 957 & 836 & 928 & 815 & 763 \\ 
 \hline
\end{tabular}
}
\label{tab:nonstoquastic_ferro_succ}
\end{center}
\end{table}

\begin{figure*}[hp]
    \centering
     \begin{minipage}[l]{0.49\textwidth}
         \includegraphics[scale=0.7]{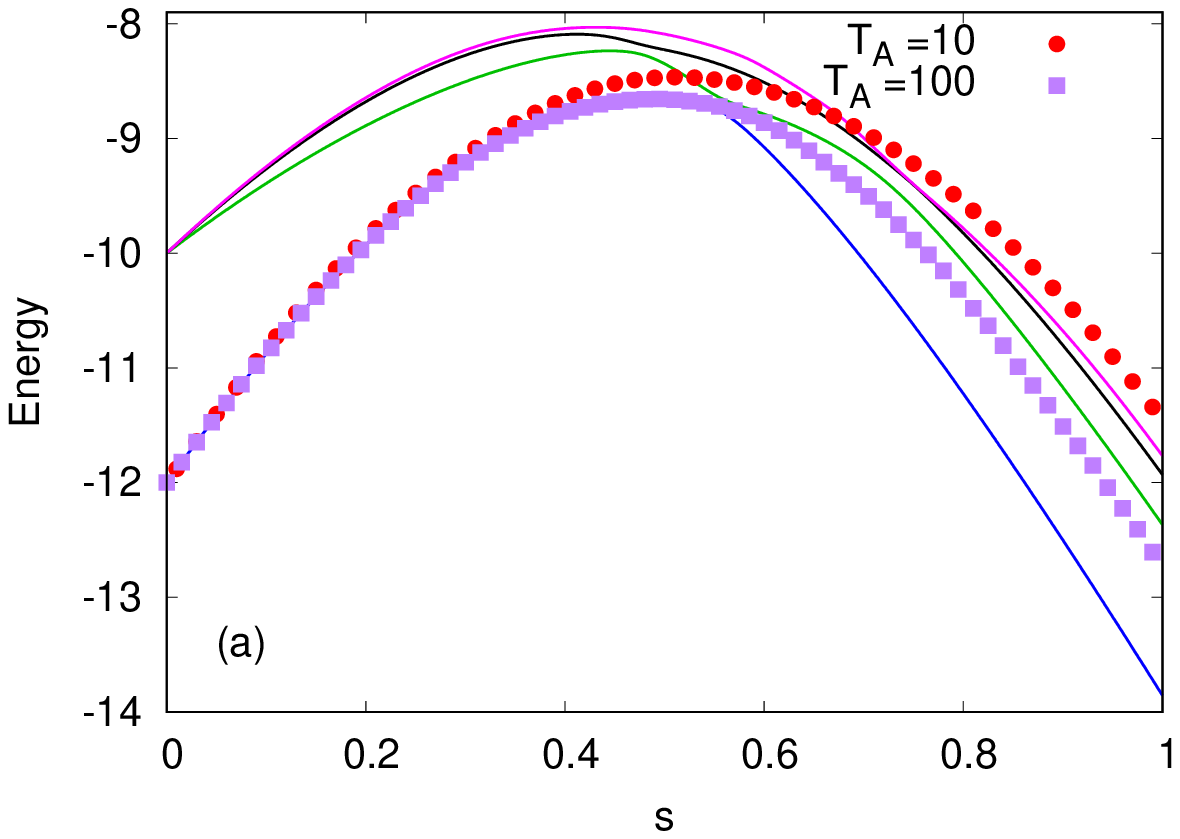}
     \end{minipage}
     \begin{minipage}[r]{0.49\textwidth}
         \centering
         \includegraphics[scale=0.7]{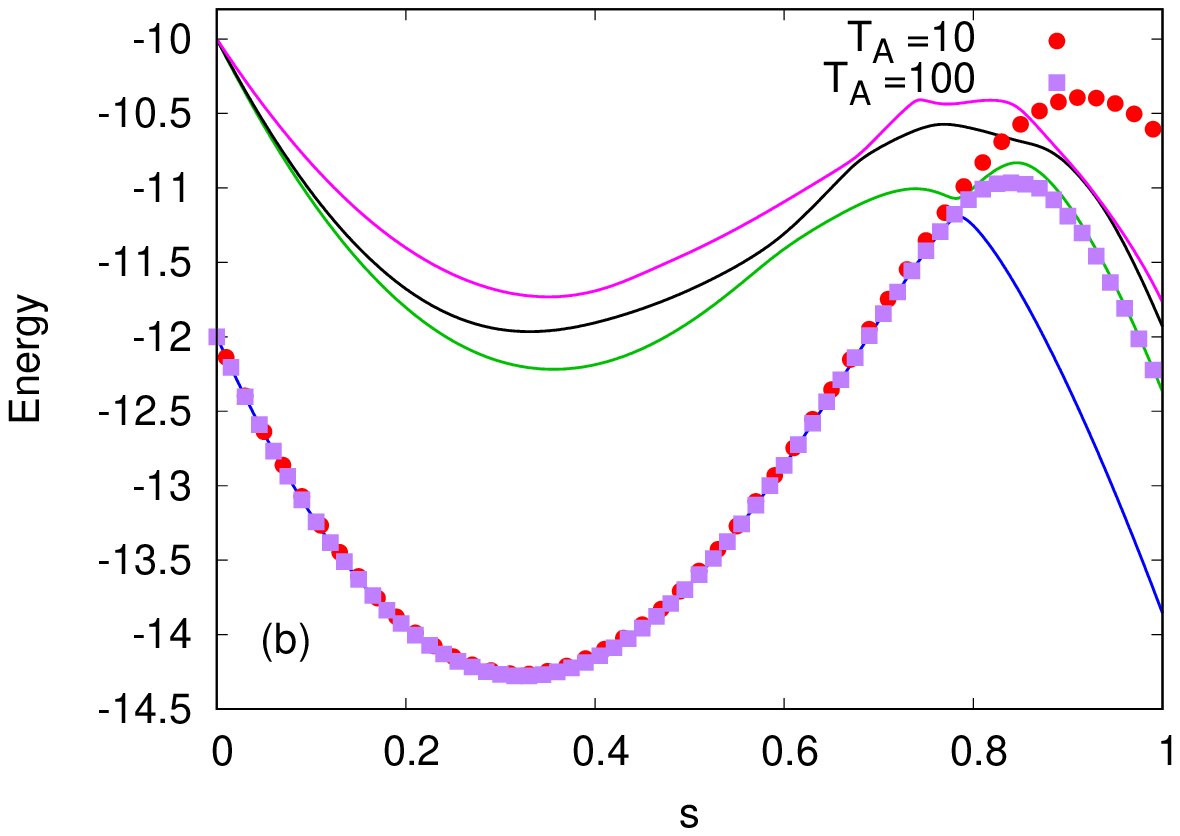}
     \end{minipage}
     \\
     \begin{minipage}[l]{0.49\textwidth}
         \includegraphics[scale=0.7]{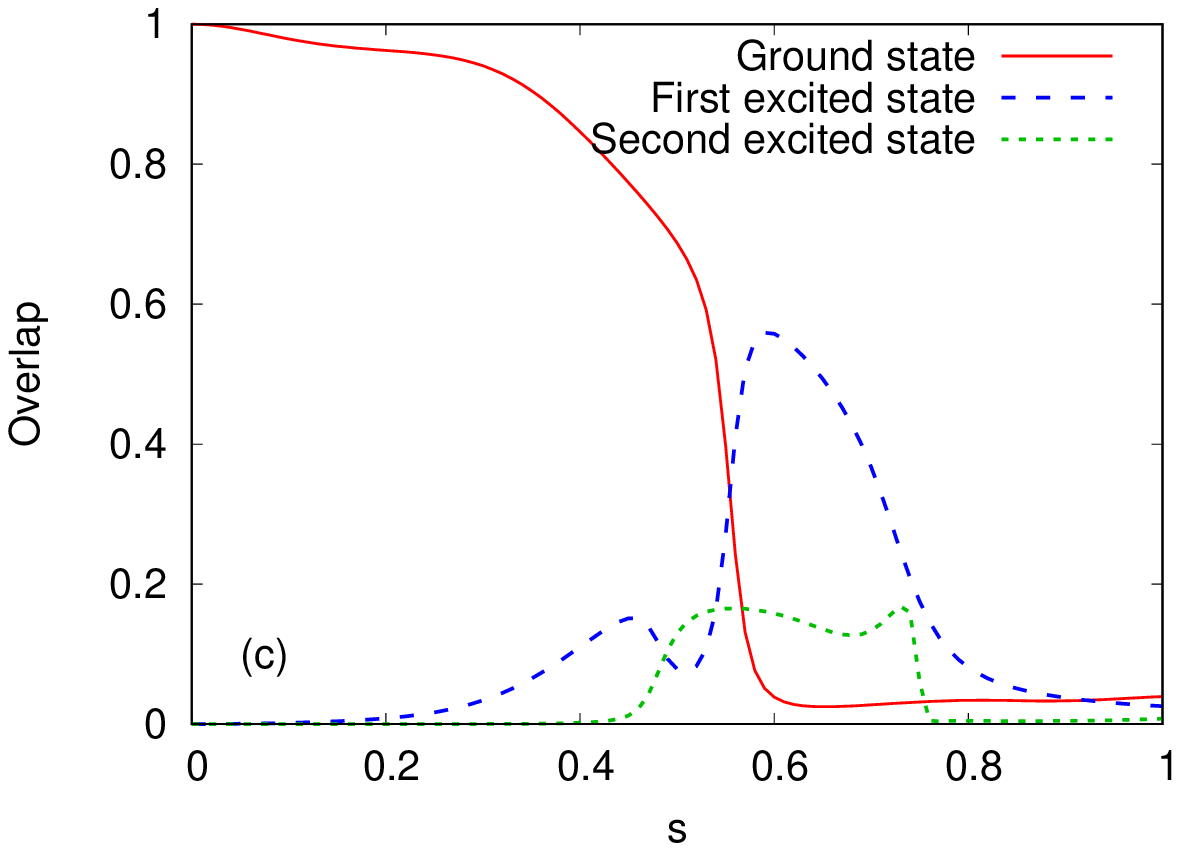}
     \end{minipage}
     \begin{minipage}[r]{0.49\textwidth}
         \centering
         \includegraphics[scale=0.7]{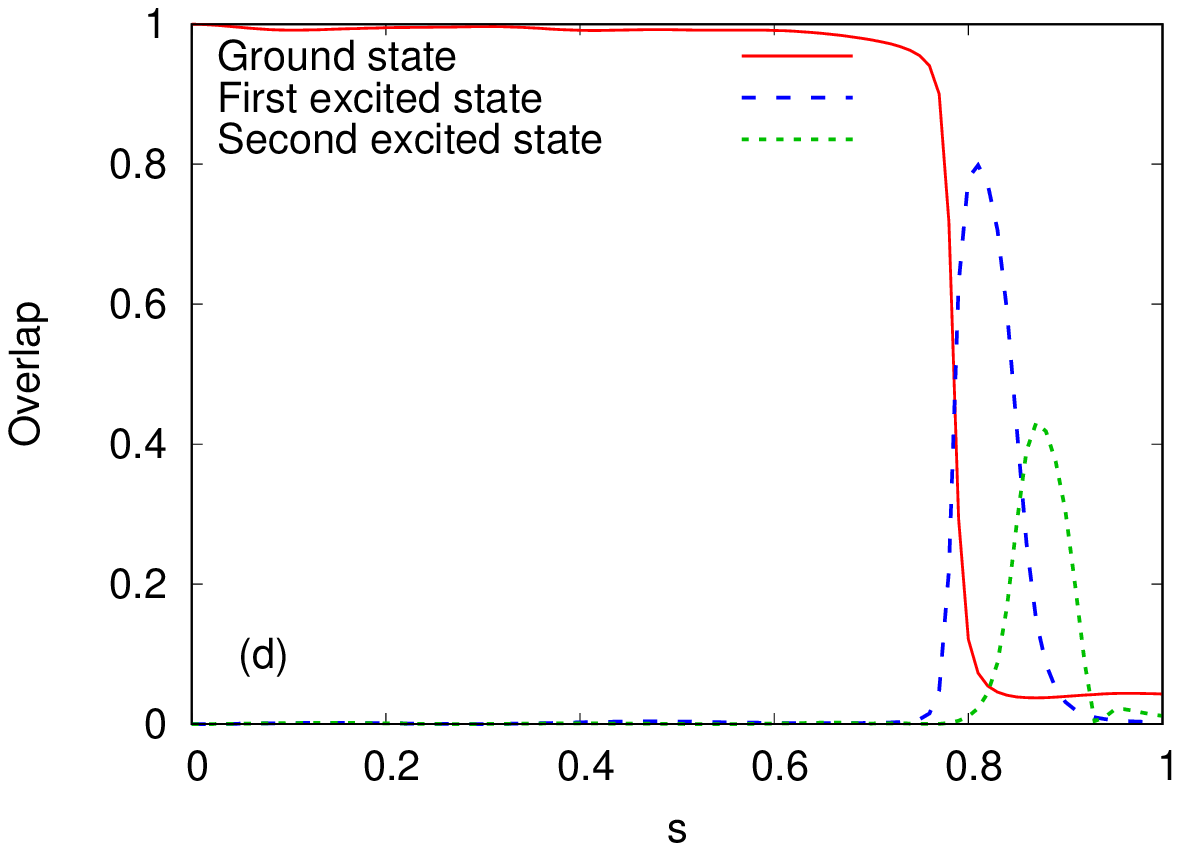}
     \end{minipage}
     \\
     \begin{minipage}[l]{0.49\textwidth}
         \includegraphics[scale=0.7]{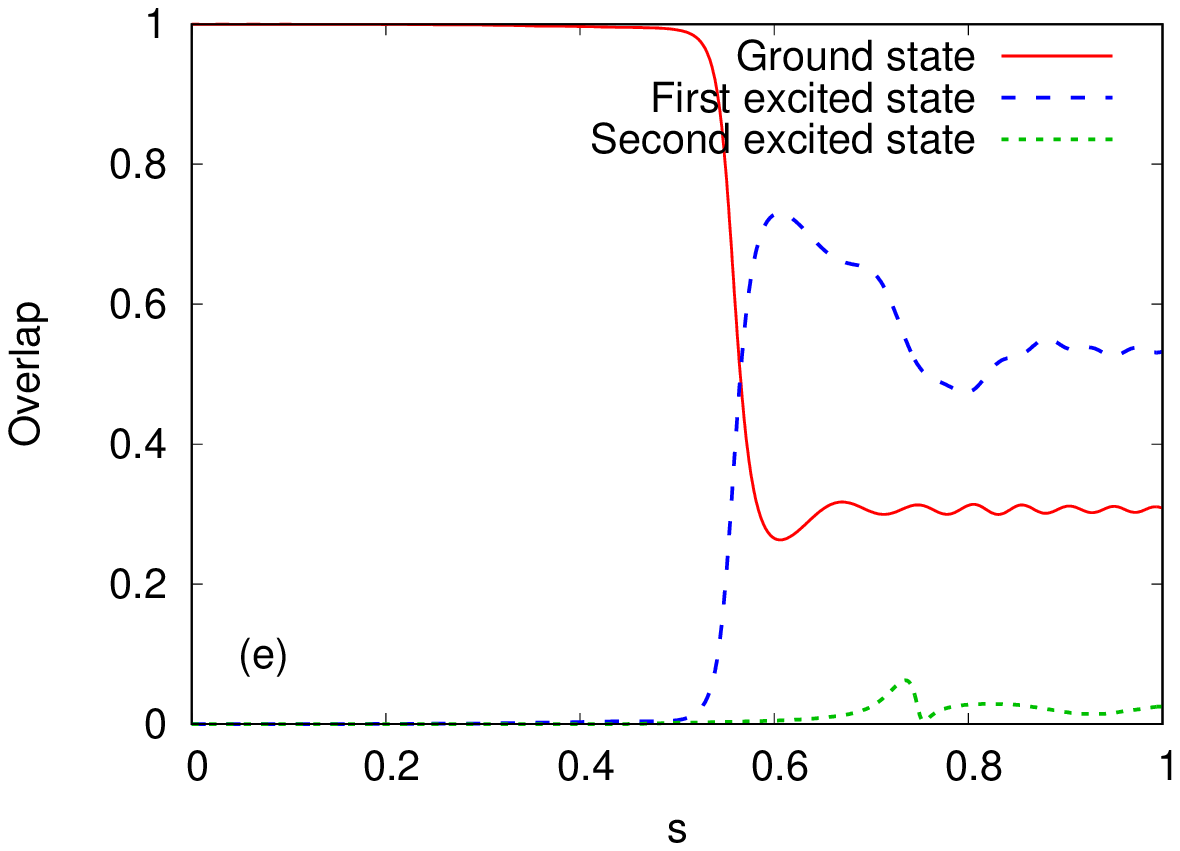}
     \end{minipage}
     \begin{minipage}[r]{0.49\textwidth}
         \centering
         \includegraphics[scale=0.7]{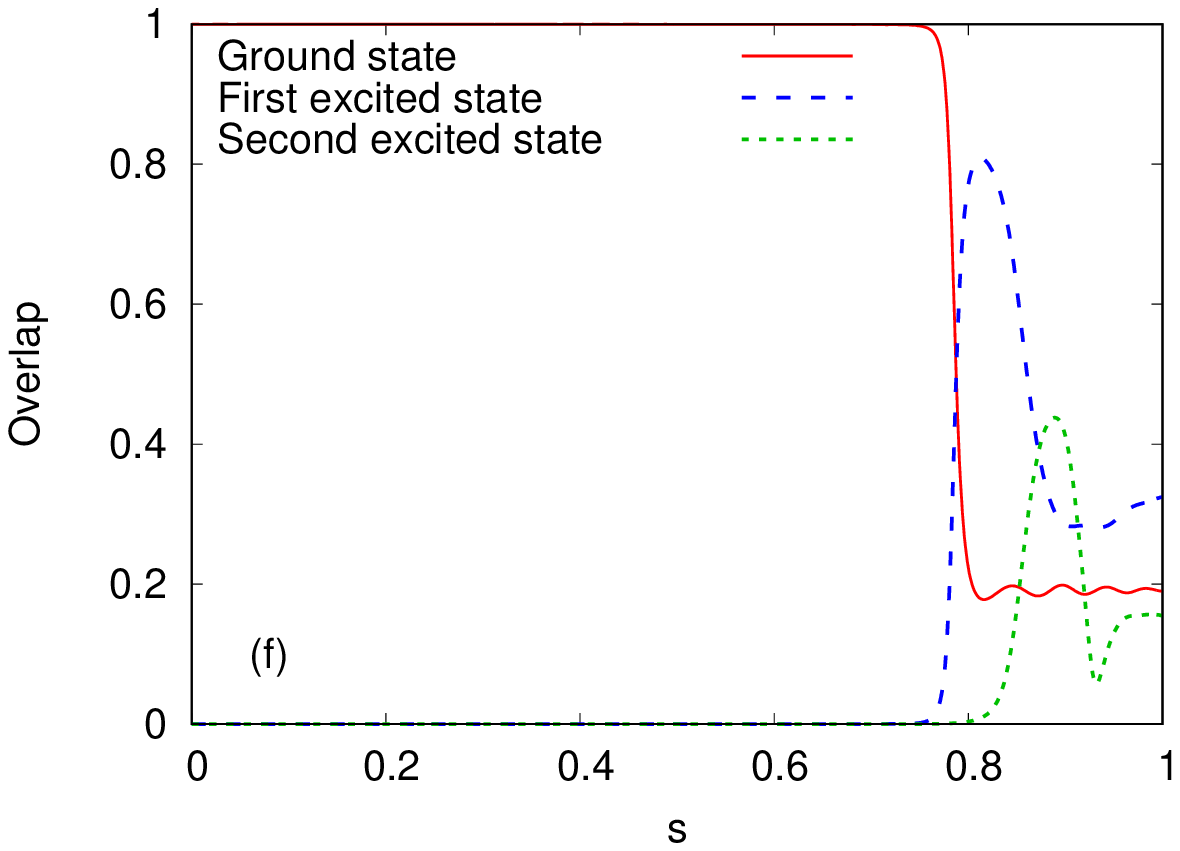}
     \end{minipage}
     \caption{(Color online) (a), (c), (e) Energy of the lowest lying states and overlap of the ground state with the lowest three energy states for quantum annealing without trigger Hamiltonian and (b), (d), (f) after adding a ferromagnetic trigger with $g$=2.0 for the 12-variable 2-SAT problem with number 99. The solid lines in panels (a) and (b) represent the lowest lying states while the points in panels (a) and (b) represent the average energy of the instantaneous state of the system for $T_A$=10, 100. The overlap is calculated for (c), (d) $T_A$=10 and (e), (f) $T_A$=100.}
     \label{fig:nonstoq_ferro_spec}
\end{figure*}

To understand the decrease in the percentage of cases with improved success probability upon increasing the annealing time, despite of an enlarged minimum energy gap, we consider one such problem. This problem is numbered as problem number ``99", and the corresponding plots are given in Fig.~\ref{fig:nonstoq_ferro_spec}.\\
\textcolor{CV}{For problem 99, the increase in the minimum energy gap, after adding the ferromagnetic trigger, is accompanied by a change in the slope of the energy of the ground state and the first-excited state as a function of the annealing parameter $s$ around the anticrossing [Fig.~\ref{fig:nonstoq_ferro_spec}(b)] in the energy spectrum. Hence, according to Eq.~(\ref{eq:adiabatic}), the success probability additionally depends on the rate of change of the Hamiltonian.\\
For the Hamiltonian without the trigger, the annealing time $T_A=10$ leads to the nonadiabatic mechanism of fast annealing, causing some of the amplitude from the ground state to shift to the first-excited state before reaching the anticrossing [Fig.~\ref{fig:nonstoq_ferro_spec}(c)]. This transferred amplitude is then lost to the second-excited state, before reaching the anticrossing between the lowest two states, thus missing the opportunity to return to the ground state. Upon reaching the anticrossing, more amplitude is transferred from the ground state to the first-excited state of the Hamiltonian, leaving an even smaller amplitude in the ground state. On the other hand, due to the enlarged minimum energy gap upon adding the ferromagnetic trigger, the state of the system deviates from the ground state only around the anticrossing [Fig.~\ref{fig:nonstoq_ferro_spec}(d)]. This results in a marginally larger overlap of the final state with the ground state in this case, compared with that for the Hamiltonian without trigger.\\
When the annealing time is increased to $T_A=100$, the state of the system for the Hamiltonian without the trigger Hamiltonian deviates from the ground state only at the anticrossing [Fig.~\ref{fig:nonstoq_ferro_spec}(e)]. However, for the Hamiltonian with the ferromagnetic trigger, the amplitude that is transferred from the ground state to the first-excited state this time, is larger owing to the shape of the resulting anticrossing, despite of its larger size [Fig.~\ref{fig:nonstoq_ferro_spec}(f)].
}

\begin{table}
\begin{center}
\caption{Number of 12-variable and 18-variable nonstoquastic problems with different numbers of anticrossings, $N_A$, after adding the ferromagnetic trigger Hamiltonian with strengths $g$, to the Hamiltonian~(\ref{eq_annealing}). Both sets consist of 1000 problems each.}
\resizebox{\columnwidth}{!}{
\begin{tabular}{ |c||c|c|c||c|c|c| } 
 \hline
 &\multicolumn{3}{c||}{12-variables} & \multicolumn{3}{c|}{18-variables}\\
 \hline
 \bm{$N_A$} & \bm{$g=0.5$} & \bm{$g=1.0$} & \bm{$g=2.0$}  & \bm{$g=0.5$} & \bm{$g=1.0$} & \bm{$g=2.0$} \\ 
 \hline
 1 & 997 & 999 & 996 & 902 & 977 & 995 \\ 
 \hline
 2 & 3 & 1 & 2 & 98 & 23 & 5 \\
 \hline
\end{tabular}
}
\label{tab:anticrossings_nonstoq_ferro}
\end{center}
\end{table}
In addition, compared with the number of problems with multiple anticrossings between the ground state and the first-excited state for the nonstoquastic problems without adding the trigger (see Table~\ref{tab:nonstoq_orig_antic}), the number after adding the ferromagnetic trigger decreases successively with increasing strength of the trigger (see Table~\ref{tab:anticrossings_nonstoq_ferro}).

\begin{figure*}[ht]
    \centering
     \begin{minipage}[c]{0.32\textwidth}
         \includegraphics[scale=0.47]{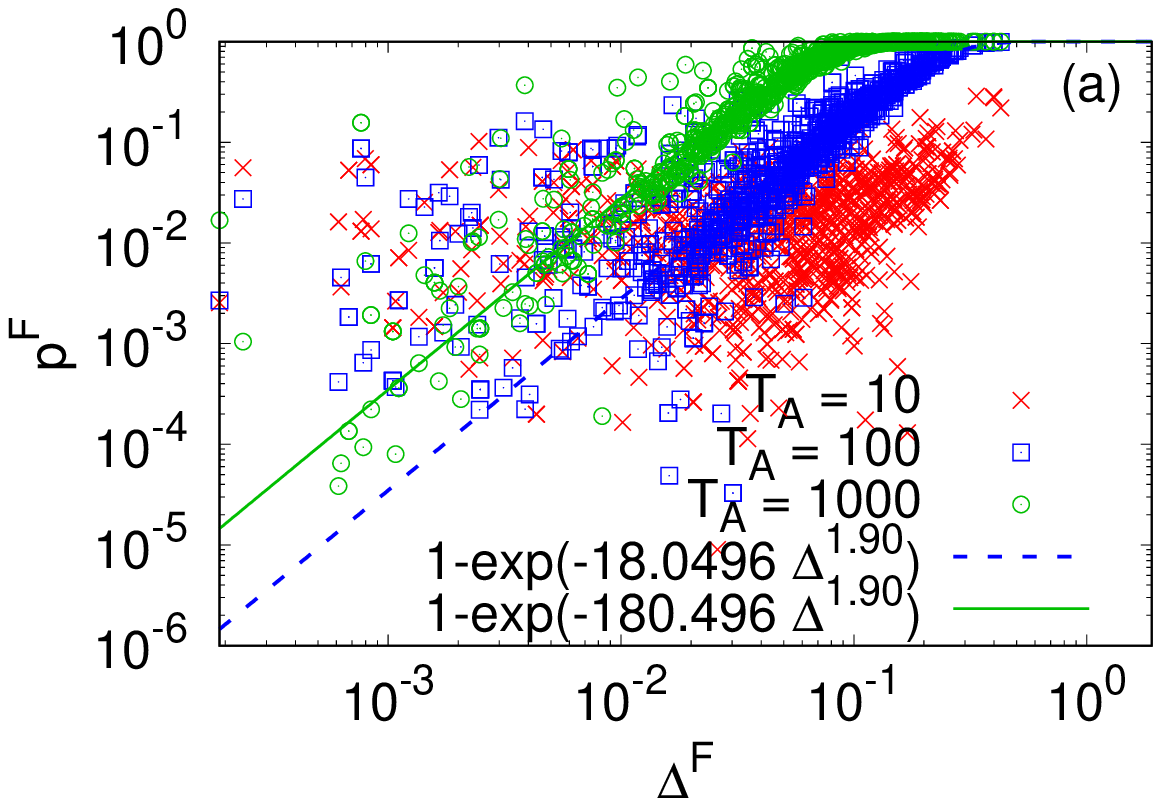}
     \end{minipage}
     \begin{minipage}[r]{0.32\textwidth}
         \includegraphics[scale=0.47]{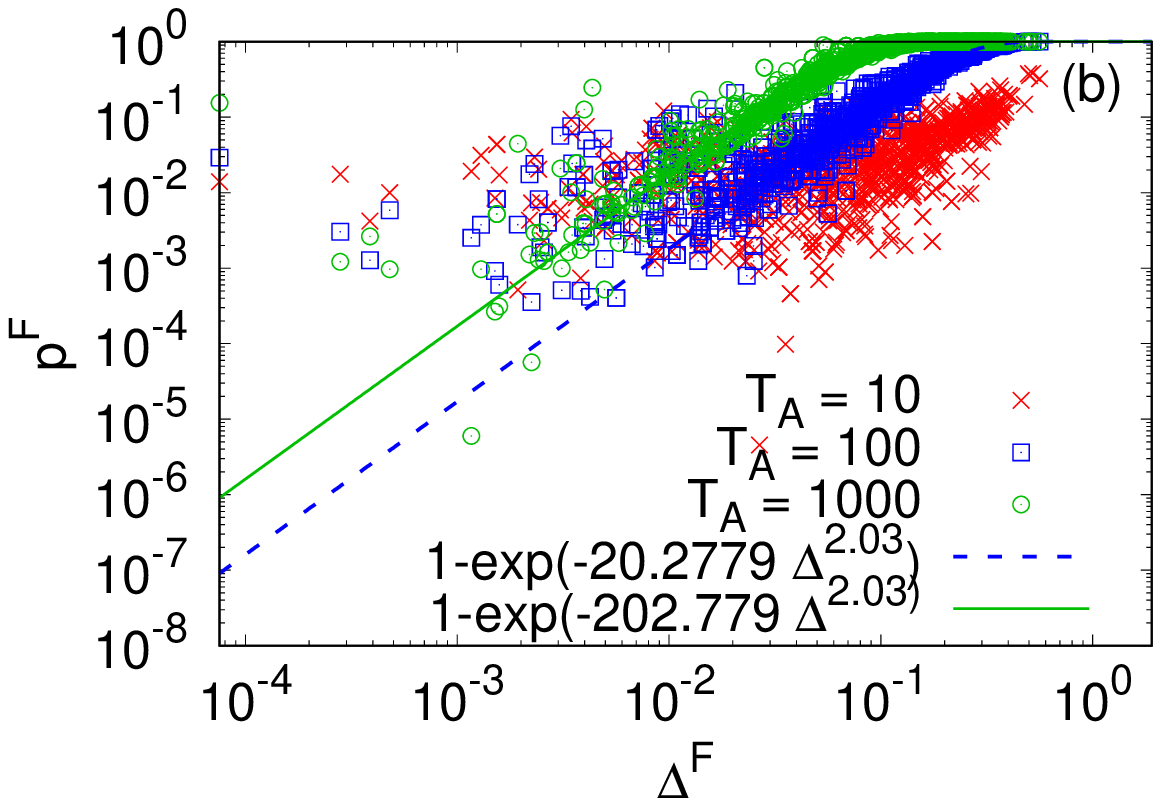}
     \end{minipage}
     \begin{minipage}[l]{0.32\textwidth}
         \includegraphics[scale=0.47]{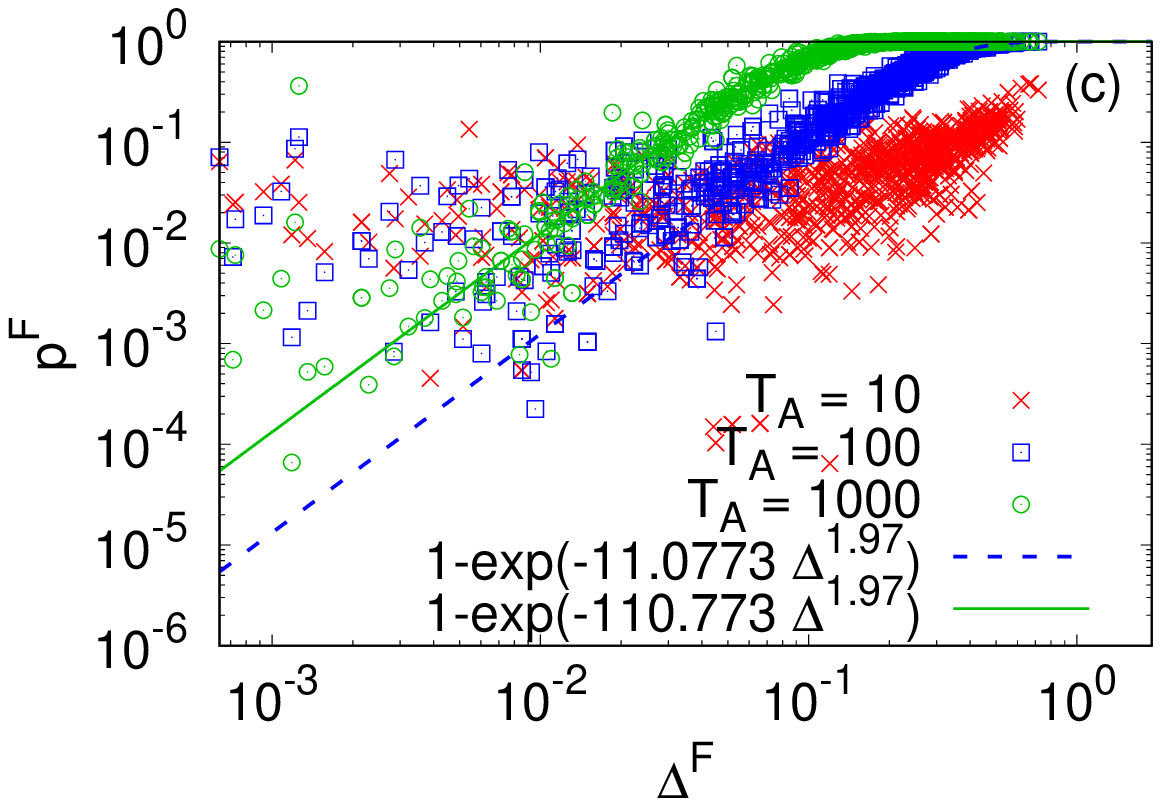}
     \end{minipage}
    \caption{(Color online) Success probability $p^F$ versus minimum energy gap $\Delta^F$ for the 18-variable nonstoquastic problems after adding the ferromagnetic trigger Hamiltonian to Hamiltonian~(\ref{eq_annealing}) with strengths (a) $g$=0.5, (b) $g$=1.0, (c) and $g$=2.0 and for various annealing times $T_A$. The lines are fits to the data.}
    \label{fig:nonstoq_ferro_succvsgap}
\end{figure*}

The success probability $p^F$ versus minimum energy gap $\Delta^F$ for the nonstoquastic problems upon adding the ferromagnetic trigger, shown in Fig.~\ref{fig:nonstoq_ferro_succvsgap}, can still be well approximated by $p=1-\exp({-a\Delta^{b}})$, with $b=1.90, 2.03, 1.97$ for $g=0.5, 1.0, 2.0$, respectively. However, the scattering of the data, especially for $T_A=10$, is larger compared with that for the 2-SAT problems (see Fig.~\ref{fig:Ferro_succvsgap}). This can be attributed to the presence of multiple anticrossings. Similarly, a comparison with the success probability versus minimum energy gaps for the nonstoquastic Hamiltonian without any triggers (see Fig.~\ref{fig:Nonstoq_Orig}) shows that the scattering in the data for $g=2.0$ is much smaller because the number of problems with multiple anticrossings has reduced (see Tables~\ref{tab:nonstoq_orig_antic} and \ref{tab:anticrossings_nonstoq_ferro}).

Finally, we add the antiferromagnetic trigger to the nonstoquastic problems. The effects of adding this trigger are very similar to those obtained for the stoquastic 2-SAT problems, i.e., the minimum energy gap reduces for a majority of the problems of the set for all chosen values of the coupling strength $g$, and the multiple anticrossings emerge between the ground state and the first-excited state of the Hamiltonian. Tables~\ref{tab:nonstoq_antiferro_mingaps} and \ref{tab:nonstoquastic_antiferro_succ} give the number of problems with an enlarged minimum energy gap and with an improved success probability, respectively.

\begin{table}
\begin{center}
\caption{Number of 12-variable and 18-variable nonstoquastic problems with enlarged minimum energy gaps after adding the antiferromagnetic trigger with strengths $g$, to the Hamiltonian~(\ref{eq_annealing}). Both sets consist of 1000 problems each.}
\begin{tabular}{ |c|c|c| } 
 \hline
 \bm{$g$} & \textbf{12-variables} & \textbf{18-variables} \\ 
 \hline
 0.5 & 331 & 406 \\ 
 \hline
 1.0 & 305 & 442 \\ 
 \hline
 2.0 & 249 & 367 \\ 
 \hline
\end{tabular}
\label{tab:nonstoq_antiferro_mingaps}
\end{center}
\end{table}

\begin{table}
\begin{center}
\caption{Number of 12-variable and 18-variable nonstoquastic problems with increased success probabilities after adding the antiferromagnetic trigger Hamiltonian with strengths $g$, to the Hamiltonian~(\ref{eq_annealing}). Both sets consist of 1000 problems each.}
\resizebox{\columnwidth}{!}{
\begin{tabular}{ |c||c|c|c||c|c|c| } 
 \hline
 &\multicolumn{3}{c||}{12-variables} & \multicolumn{3}{c|}{18-variables}\\
 \hline
 \textbf{$g$} & \bm{$T_A=10$} & \bm{$T_A=100$} & \bm{$T_A=1000$}  & \bm{$T_A=10$} & \bm{$T_A=100$} & \bm{$T_A=1000$} \\ 
 \hline
 0.5 & 323 & 353 & 356 & 488 & 444 & 467 \\ 
 \hline
 1.0 & 240 & 353 & 344 & 393 & 457 & 481 \\ 
 \hline
 2.0 & 5 & 150 & 238 & 8 & 240 & 380 \\ 
 \hline
\end{tabular}
}
\label{tab:nonstoquastic_antiferro_succ}
\end{center}
\end{table}

\begin{figure*}[ht]
    \centering
     \begin{minipage}[c]{0.32\textwidth}
         \includegraphics[scale=0.47]{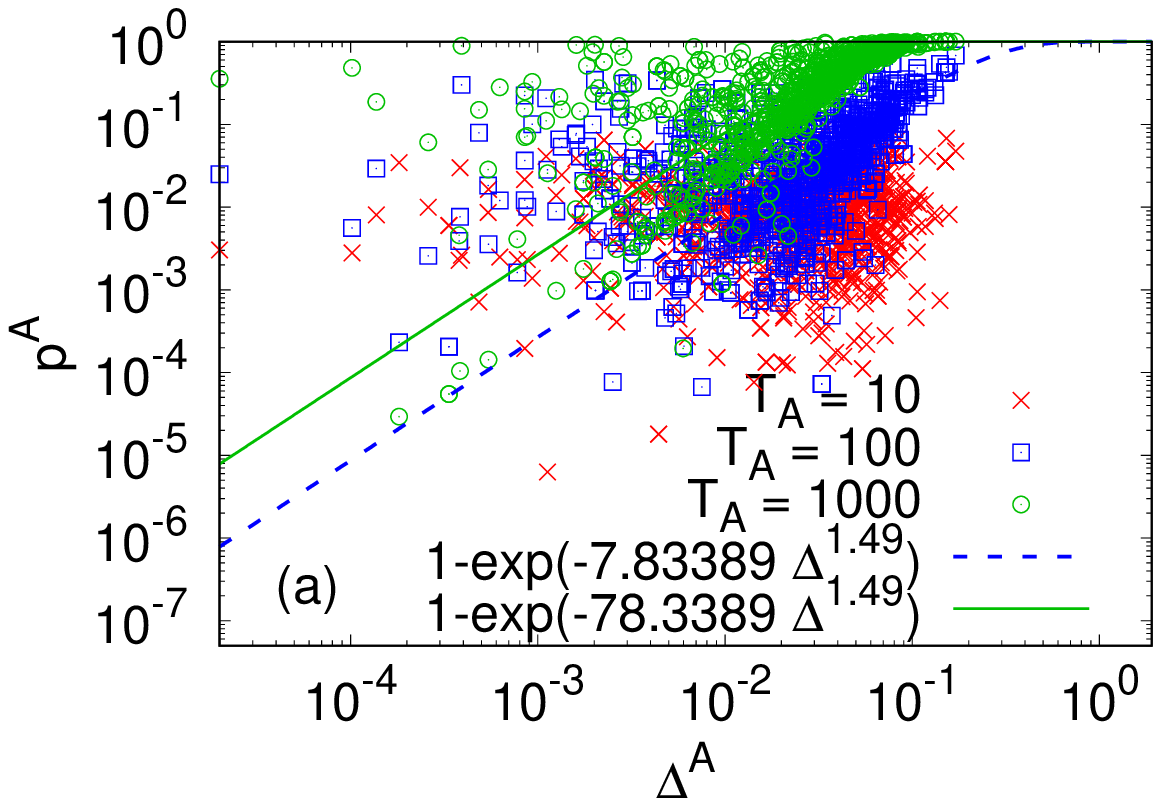}
     \end{minipage}
     \begin{minipage}[r]{0.32\textwidth}
         \includegraphics[scale=0.47]{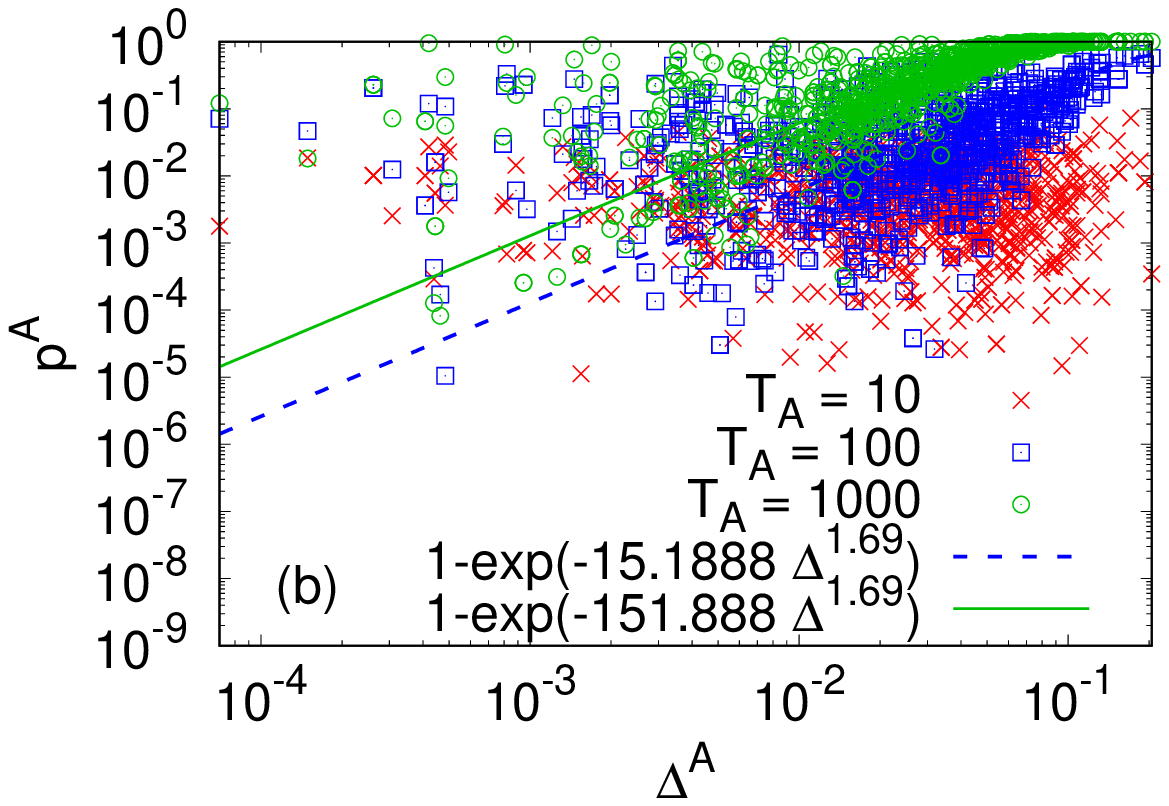}
     \end{minipage}
     \begin{minipage}[r]{0.3\textwidth}
         \includegraphics[scale=0.47]{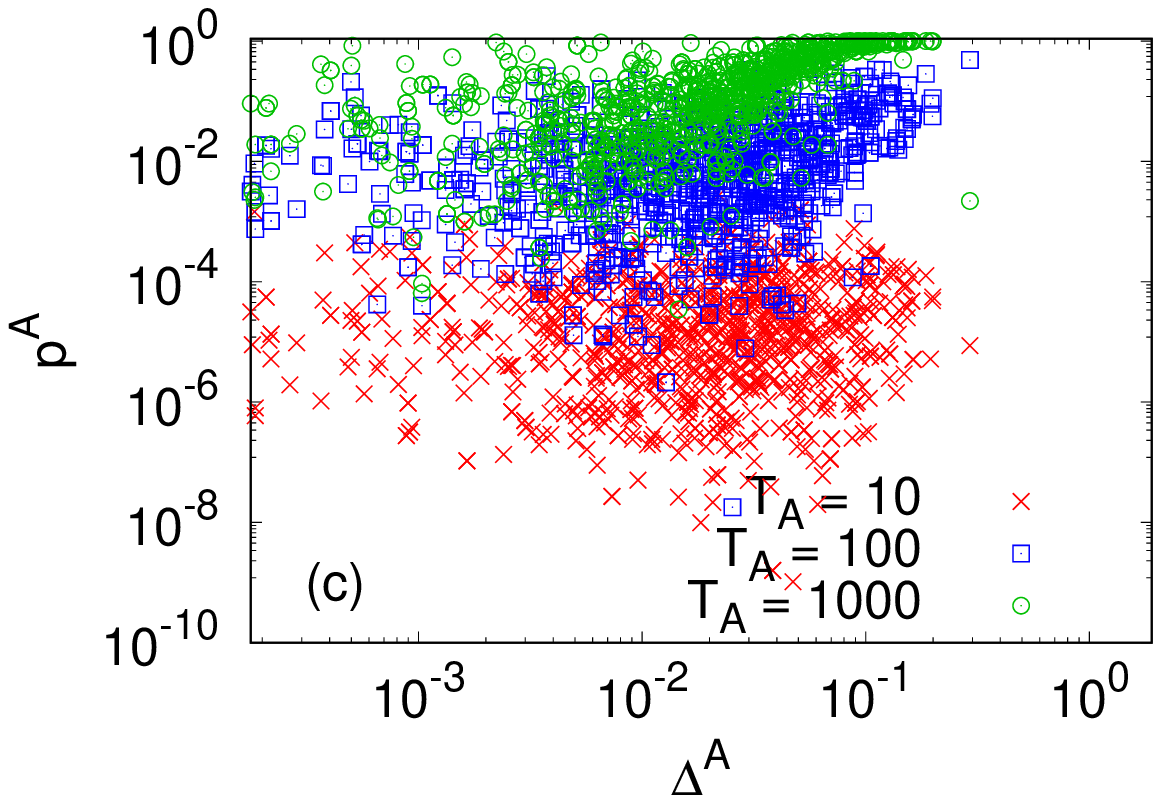}
     \end{minipage}
    \caption{(Color online) Success probability $p^A$ versus minimum energy gap $\Delta^A$ for the 18-variable nonstoquastic problems after adding the antiferromagnetic trigger Hamiltonian to Hamiltonian~(\ref{eq_annealing}) with strengths (a) $g$=0.5, (b) $g$=1.0, and (c) $g$=2.0 and for various annealing times $T_A$. The lines are fits to the data.}
    \label{fig:nontoq_antiferro_succvsgap}
\end{figure*}

Table~\ref{tab:anticrossings_nonstoq} shows the number of problems with different number of anticrossings after adding the antiferromagnetic trigger with coupling strengths $g=0.5, 1.0, 2.0$. It can be noted that compared with the 2-SAT problems (see Table~\ref{tab:anticrossings}), the nonstoquastic problems have a larger number of problems with more anticrossings. This explains why more scattering is seen in the scatter plot of the success probability $p^A$ versus minimum energy gap $\Delta^A$, as shown in Fig.~\ref{fig:nontoq_antiferro_succvsgap}.

\begin{table}
\begin{center}
\caption{Number of 12-variable and 18-variable nonstoquastic problems with different number of anticrossings, $N_A$, after adding the antiferromagnetic trigger Hamiltonian with strengths $g$, to the Hamiltonian~(\ref{eq_annealing}). Both sets consist of 1000 problems each.}
\resizebox{\columnwidth}{!}{
\begin{tabular}{ |c||c|c|c||c|c|c| } 
 \hline
 &\multicolumn{3}{c||}{12-variables} & \multicolumn{3}{c|}{18-variables}\\
 \hline
 \bm{$N_A$} & \bm{$g=0.5$} & \bm{$g=1.0$} & \bm{$g=2.0$}  & \bm{$g=0.5$} & \bm{$g=1.0$} & \bm{$g=2.0$} \\ 
 \hline
 1 & 671 & 141 & 4 & 276 & 18 & 0 \\ 
 \hline
 2 & 323 & 620 & 88 & 597 & 306 & 25 \\ 
 \hline
 3 & 6 & 231 & 410 & 121 & 494 & 166 \\ 
 \hline
 4 & 0 & 8 & 396 & 6 & 171 & 407 \\ 
 \hline
 5 & 0 & 0 & 101 & 0 & 11 & 310 \\ 
 \hline
 6 & 0 & 0 & 1 & 0 & 0 & 86 \\ 
 \hline
 7 & 0 & 0 & 0 & 0 & 0 & 6 \\ 
 \hline
\end{tabular}
}
\label{tab:anticrossings_nonstoq}
\end{center}
\end{table}

\section{Conclusion}
\label{conclusion}

In this work, we studied quantum annealing for two large sets of problems, namely, 2-SAT problems and nonstoquastic problems. A third Hamiltonian, namely, a trigger Hamiltonian, was added to the conventional transverse Ising spin Hamiltonian, which describes the quantum annealing process. This trigger could be of ferromagnetic or antiferromagnetic type. \\

Since for a majority of the problems studied, for a sufficiently long annealing time the Hamiltonian can be approximated by the one of a two-level system, the success probabilities for these problems can be explained on the basis of the Landau-Zener model.\\
For the 2-SAT problems, the addition of the ferromagnetic trigger Hamiltonian to the Hamiltonian~(\ref{eq_annealing}) of quantum annealing, always leads to an enlargement of the minimum energy gap occurring between the ground state and the first-excited state of the Hamiltonian, and thus, an enhancement in the success probability. This enhancement of the success probability increases with increasing the strength of the trigger.\\
However, upon adding an antiferromagnetic trigger Hamiltonian, the minimum energy gap can either increase or decrease. Moreover, adding the antiferromagnetic trigger leads to a distortion in the energy spectrum of the Hamiltonian, most important of which is the emergence of multiple anticrossings. Overall, this leads to a decrease in the success probability upon adding the antiferromagnetic trigger, for a majority of the problems studied, but in some cases it can lead to an increase of the success probability, depending on the nonadiabatic processes that take place.

Thus, adding the ferromagnetic trigger Hamiltonian to the Hamiltonian~(\ref{eq_annealing}) guarantees an improvement in the performance of quantum annealing for solving the studied 2-SAT problems. On the other hand, adding the antiferromagnetic trigger can lead to larger enhancements of the success probability owing to the nonadiabatic mechanisms (i.e., an even number of comparably small anticrossings, fast annealing, or stretching of the anticrossing), but this improvement is hard to predict systematically because there is no prior information about the energy spectrum of the Hamiltonian.

For the nonstoquastic problems, the differences in the resulting effects of adding the two triggers on the success probability are not as distinct. For a majority of the problems, adding the ferromagnetic trigger Hamiltonian leads to an increase in the minimum energy gap. However, increasing the annealing time can result in a decrease in the number of problems with improved success probability, as the change in the slope of the energy of the ground state and the first-excited state (around the anticrossing) in the energy spectrum also becomes relevant. The effects of adding the antiferromagnetic trigger Hamiltonian, on the other hand, are similar to the effects seen for the 2-SAT problems.

\section{Acknowledgements}

The authors gratefully acknowledge the Gauss Centre for Supercomputing e.V. (www.gauss-centre.eu) for funding this project by providing computing time on the GCS Supercomputer JUWELS at J\"{u}lich Supercomputing Centre (JSC). The authors also gratefully acknowledge the computing time granted through JARA on the supercomputer JURECA at Forschungszentrum J\"{u}lich. V.M. acknowledges support from the project JUNIQ that has received funding from the German Federal Ministry of Education and Research (BMBF) and the Ministry of Culture and Science of the State of North Rhine-Westphalia.

\appendix
\label{appendix}
\section{2-spin system}
\label{sec:appendix_A}
\begin{figure*}[ht]
    \centering
     \begin{minipage}[c]{0.32\textwidth}
         \includegraphics[scale=0.47]{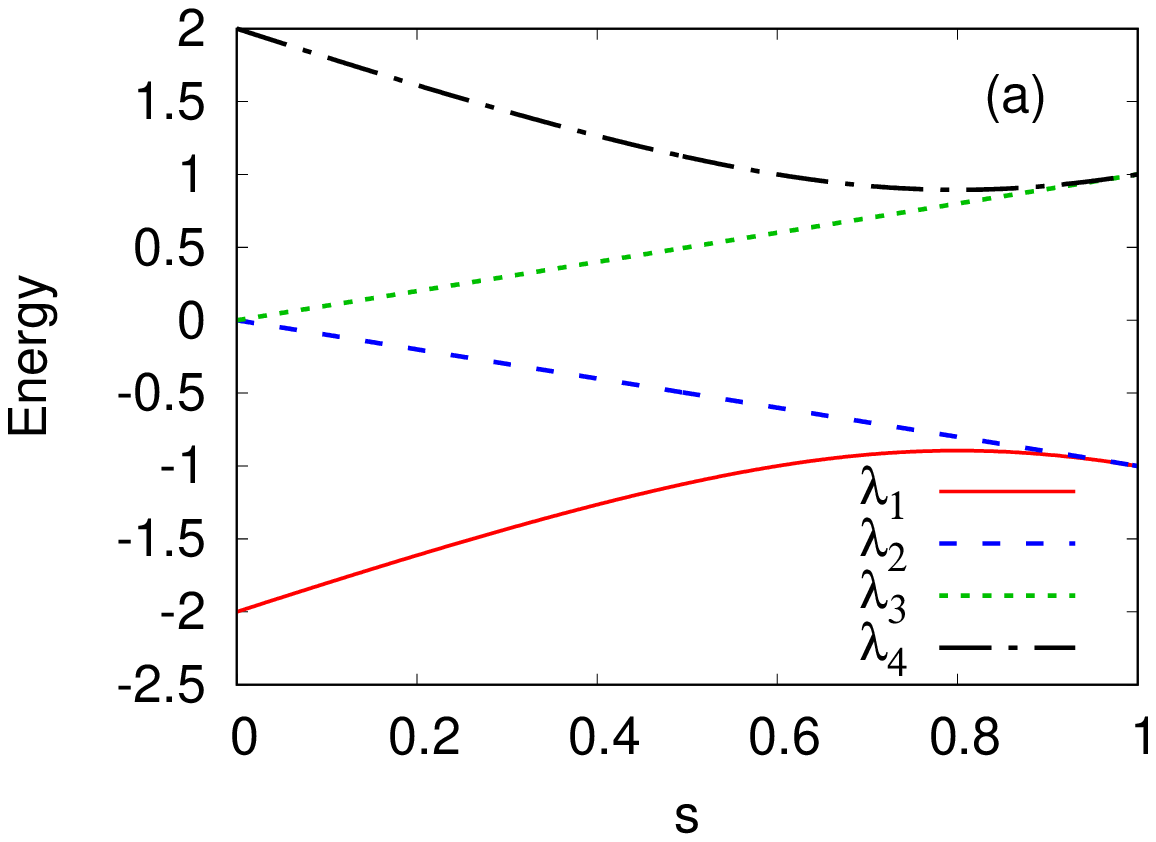}
     \end{minipage}
     \begin{minipage}[r]{0.32\textwidth}
         \includegraphics[scale=0.47]{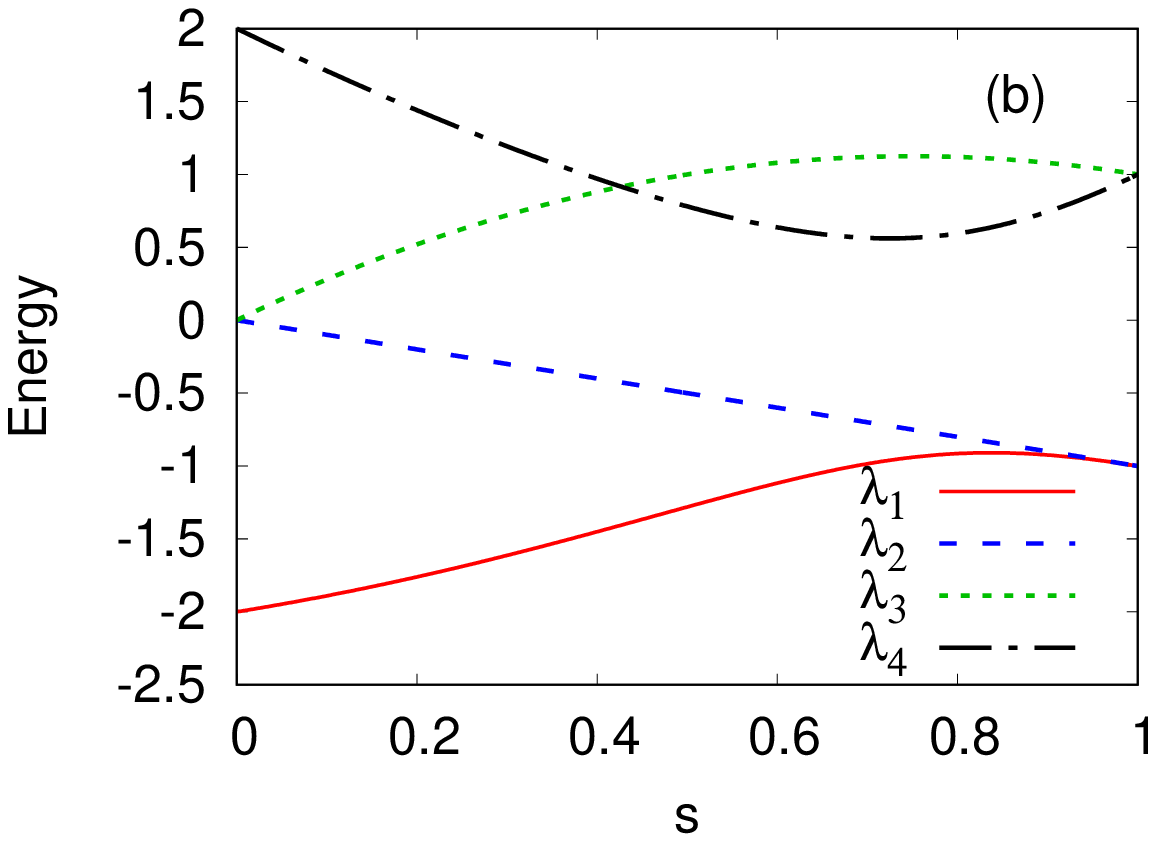}
     \end{minipage}
     \begin{minipage}[r]{0.32\textwidth}
         \includegraphics[scale=0.47]{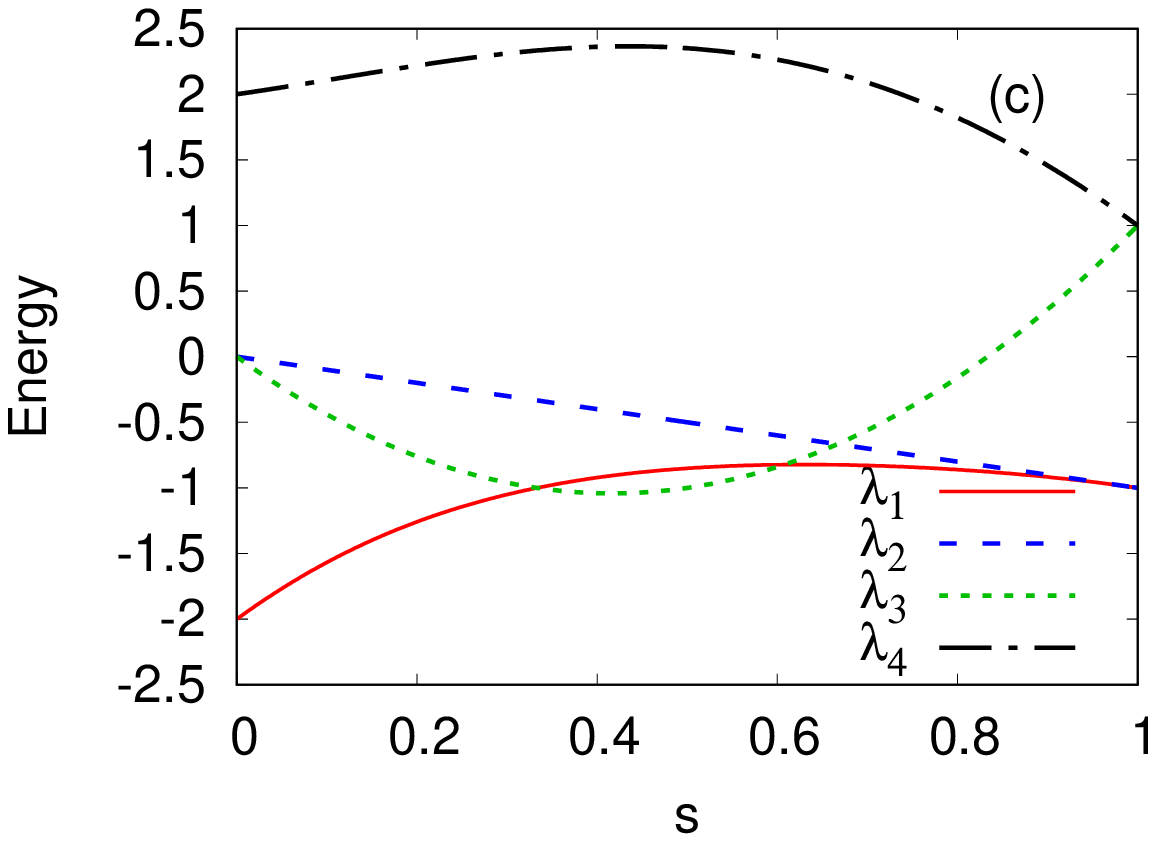}
     \end{minipage}\\
    \caption{(Color online)  Exact energy spectra of the 2-spin model  for the case without trigger Hamiltonian being added to Hamiltonian~(\ref{eq_annealing}) (a), with a ferromagnetic trigger Hamiltonian with $g$=1.0 being added (b) and with an antiferromagnetic trigger Hamiltonian with $g$=3.0 being added (c).}
    \label{fig:model}
\end{figure*}

We study a system consisting of only two spins to investigate if it can show similar modifications in the energy spectrum when adding the trigger Hamiltonian to the Hamiltonian~(\ref{eq_annealing}) as the problems considered in the main body of this work.
To make the trigger Hamiltonian more general, a coupling with strength  $J_y$ is added, so that
\begin{equation}
    H_T = g(J_{1,2}^x \sigma_1^x \sigma_2^x + J_{1,2}^y\sigma_1^y \sigma_2^y).
\end{equation}
The energy spectrum is then determined for the full Hamiltonian given in Eq.(\ref{eq:totalHamil}). For simplicity, the bias terms ($h_z$) in the problem Hamiltonian are set to 0. The eigenvalues of the full Hamiltonian are given by
\begin{equation}
\begin{aligned}
    \lambda_1 &= s^2gJ_x - sgJ_x - R \\
    \lambda_2 &= -s^2gJ_x + s^2gJ_y + sgJ_x - sgJ_y - sJ_z\\
    \lambda_3 &= -s^2gJ_x - s^2gJ_y + sgJ_x + sgJ_y + sJ_z\\
    \lambda_4 &= s^2gJ_x - sgJ_x + R, 
\end{aligned}
\label{eq:eigenvalues}
\end{equation}
where $R=$ $(s^4g^2J_y^2 - 2s^3g^2J_y^2 + 2s^3gJ_yJ_z + s^2g^2J_y^2- 2s^2gJ_yJ_z$ $ + s^2J_z^2 + 2s^2 - 4s + 2)^{1/2}$.\\
With $J_z$ chosen to be 1, Fig.~\ref{fig:model}(a) shows the energy spectrum without the trigger Hamiltonian being added to Hamiltonian~(\ref{eq_annealing}), while Figs.~\ref{fig:model}(b) and Fig.~\ref{fig:model}(c) show the spectra when adding the ferromagnetic ($J_x=J_y=1$) and antiferromagnetic trigger Hamiltonians ($J_x=J_y=-1$), respectively. Interestingly, for this smaller problem, adding the antiferromagnetic trigger with a large trigger strength, leads to the creation of two crossings between the instantaneous ground and first-excited states, and another crossing between the first- and second-excited states.\\
Close to $s=1$, the energy gap between the ground state and the first-excited state is given by $|\lambda_2-\lambda_1|$ for all the three cases. Similarly, if $J_z<0$, the energy gap is given by $|\lambda_4-\lambda_3|$. From Eq.~(\ref{eq:eigenvalues}), it is seen that the energy gap between these states up to leading orders in $s$, is given by
\begin{equation}
\begin{aligned}
    \Delta_{1,2}&=2(1-s)\bigg( \frac{(1-s)}{2sJ_z} + sgJ_x - sgJ_Y \bigg),\\
    \Delta_{3,4}&=2(1-s)\bigg(-\frac{(1-s)}{2sJ_z} + sgJ_x + sgJ_Y\bigg).
\end{aligned}
\label{eq:gap}
\end{equation}
This result has a similar form as the one presented in Ref.~\cite{lykiardopoulou2020improving}. According to Eq.~(\ref{eq:gap}), while $J_x$ has similar effects on gaps $\Delta_{1,2}$ and $\Delta_{3,4}$, the $J_y$ term has opposite signs in both formulas. However, since the ground state of the problem Hamiltonian is degenerate in this case, the relevant energy gap is no longer the one between the ground state and the first-excited state, but the one between the first and the second-excited states of the Hamiltonian.\\

The degeneracy of the ground state of the problem Hamiltonian can be lifted by adding the bias terms to the problem. However, the analytical formula for the eigenvalues becomes lengthy, and is omitted here. Lastly, since the whole spectrum cannot be well described in the limit $s=1$, the model cannot explain all the alterations in the energy spectrum.

\section{Suzuki Trotter Product Formula}
\label{sec:appendix_B}
For solving the time-dependent Schr{\"o}dinger equation, the unitary matrix exponential, $U(t)$, given by

\begin{equation}
    U(t) = e^{-itH} = e^{it(H_1+....+H_K)}, 
\end{equation}
needs to be computed. The Lie-Suzuki-Trotter product formula  \cite{trotter1959product,suzuki1977monte} can be utilized to construct unitary approximations to the evolution operator, such that,
\begin{equation}
    U(t)= {\lim_{m \to \infty}} \prod_{k=1}^{K} \left(e^{-itH_k/m}\right)^m.
\end{equation}

Defining $\tau=t/m$, as the time step over which the evolution operator is applied, the first-order approximation to $U(t)$ is given by \cite{de1987product}
\begin{equation}
    \widetilde{U_1}(\tau) = e^{-i \tau H_1}...e^{-i \tau H_K}, 
\end{equation}
for sufficiently small $\tau$, such that $\tau ||H|| \ll 1$. To improve the accuracy, a second-order approximation to U(t) is constructed from $\widetilde{U_1(t)}$
\begin{equation}
\begin{split}
    \widetilde{U_2}(\tau) &= \tilde{U_1}^\dagger(-\tau/2) \tilde{U_1}(\tau/2)\\
    &= e^{-i \tau H_K/2}...e^{-i \tau H_1/2}e^{-i \tau H_1/2}...e^{-i \tau H_K/2}.
\end{split}
\end{equation}
The error, calculated using the second norm of the difference between $U(\tau)$ and $\widetilde{U_2}(\tau)$, is given by \cite{de1987product}
\begin{equation}
    ||U(\tau) - \widetilde{U_2}(\tau)|| \leq c \tau^3,
\end{equation}
for a positive constant $c$. Since the whole annealing process requires $m$ such time steps \cite{de1987product}, 
\begin{equation}
    ||U(t) - \widetilde{U_2}(m \tau)|| \leq mc \tau^3 = ct \tau^2 ,
\end{equation}
since $m \tau =t$.

\textcolor{CV}{
\section{Individual problem Hamiltonians}
\label{sec:appendix_D}
In Table~\ref{tab:problems}, we provide the magnetic fields and the couplings constituting the individual problem Hamiltonians discussed in this work.
\begin{table}
\begin{center}
\caption{Magnetic fields and couplings constituting the individual 12-spin problems discussed in the paper. The top row of the table corresponds to the labels of the problems in the set.}
\resizebox{\columnwidth}{!}{
\color{CV}\begin{tabular}{ |c|c|c|c||c|c|c|c||c|c|c|c||c|c|c|c| }
 \hline
 \multicolumn{4}{|c||}{709} & \multicolumn{4}{c||}{950} & \multicolumn{4}{c||}{103} & \multicolumn{4}{c|}{99}\\ 
 \hline
 \bm{$i$} & \bm{$h_i^z$} & \bm{$i,j$} & \bm{$J_{i,j}^z$}& \bm{$i$} & \bm{$h_i^z$} & \bm{$i,j$} & \bm{$J_{i,j}^z$} & \bm{$i$} & \bm{$h_i^z$} & \bm{$i,j$} & \bm{$J_{i,j}^z$} & \bm{$i$} & \bm{$h_i^z$} & \bm{$i,j$} & \bm{$J_{i,j}^z$} \\
\hline 
1   & 0   & 1,4   & 1   & 1   & 1   & 1,11   & -1   & 1   & -1   & 1,4   & 1   & 1   & 1   & 1,9   & 1\\  
\hline 
2   & 1   & 1,11   & 1   & 2   & 4   & 2,4   & -1   & 2   & 1   & 1,10   & -1   & 2   & 1   & 2,9   & 1\\  
\hline 
3   & 0   & 2,7   & 1   & 3   & -1   & 2,5   & 1   & 3   & -1   & 1,11   & -1   & 3   & -1   & 3,4   & -1\\  
\hline 
4   & 0   & 3,7   & -1   & 4   & 1   & 2,7   & 1   & 4   & 1   & 2,8   & 1   & 4   & 0   & 4,11   & 1\\  
\hline 
5   & -1   & 3,11   & -1   & 5   & -1   & 2,8   & 1   & 5   & -1   & 3,7   & -1   & 5   & 0   & 5,6   & 1\\  
\hline 
6   & 0   & 4,9   & -1   & 6   & -1   & 2,9   & 1   & 6   & -1   & 3,8   & 1   & 6   & 1   & 5,8   & -1\\  
\hline 
7   & -4   & 5,7   & -1   & 7   & -1   & 2,11   & 1   & 7   & -1   & 3,9   & -1   & 7   & 1   & 7,9   & 1\\  
\hline 
8   & -1   & 6,7   & -1   & 8   & -1   & 3,11   & 1   & 8   & 0   & 5,11   & 1   & 8   & 0   & 7,10   & 0\\  
\hline 
9   & -1   & 6,10   & 1   & 9   & 1   & 6,9   & 1   & 9   & 1   & 6,11   & 1   & 9   & -2   & 8,12   & 1\\  
\hline 
10   & -1   & 7,8   & -1   & 10   & 1   & 9,12   & -1   & 10   & -1   & 9,11   & 1   & 10   & -1   & 9,10   & 1\\  
\hline 
11   & 1   & 7,12   & 1   & 11   & 2   & 10,11   & -1   & 11   & 2   & 9,12   & 1   & 11   & 0   & 10,11   & 1\\  
\hline 
12   & 0   & 9,11   & 1   & 12   & 1   & 10,12   & 0   & 12   & -1   & 1,2   & 0   & 12   & 0   & 10,12   & -1\\  
\hline 
   &    & 9,12   & -1   &    &    & 1,2   & 0   &    &    & 1,3   & 0   &    &    & 1,2   & 0\\  
\hline
\end{tabular}
}
\label{tab:problems}
\end{center}
\end{table}
}

\section{Results for quantum annealing for 12-variable nonstoquastic problems}
\label{sec:appendix_C}
\begin{figure}
    \centering
    \includegraphics[scale=0.7]{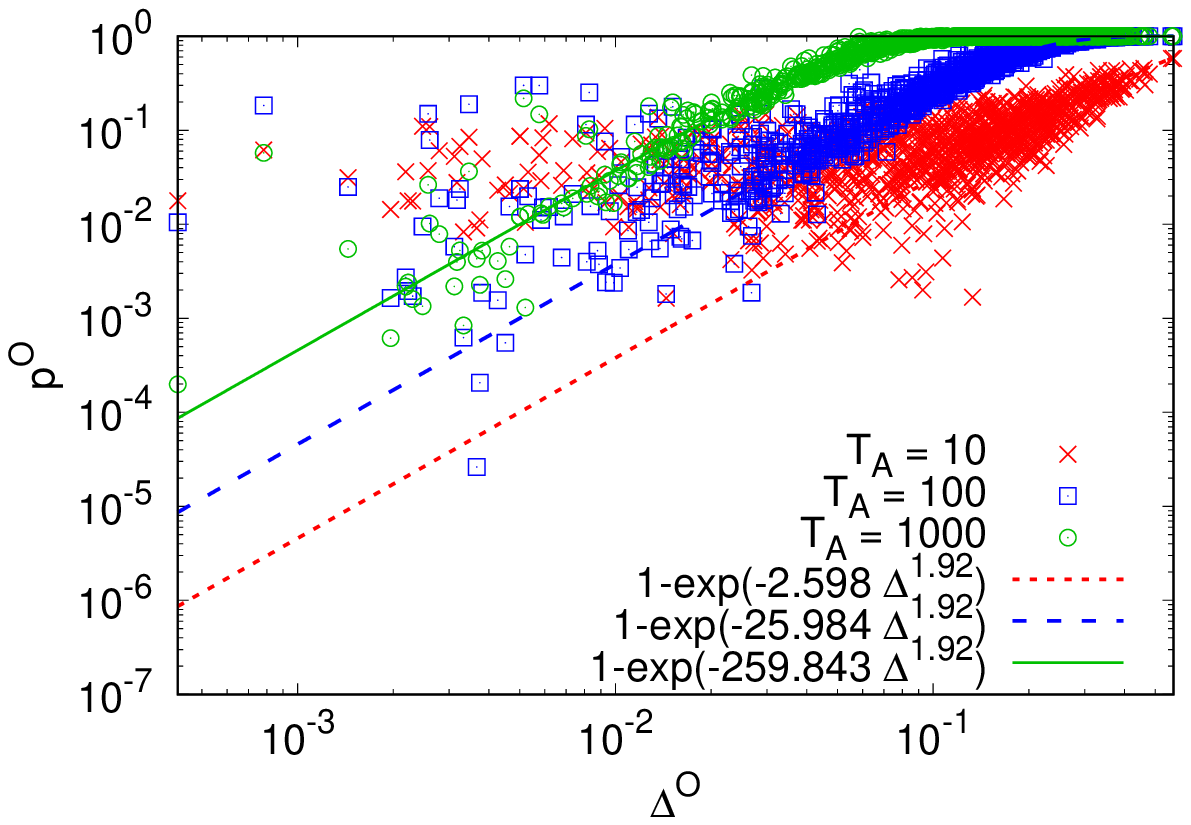}
    \caption{(Color online) Success probability $p^O$ versus minimum energy gap $\Delta^O$ for the 12-variable nonstoquastic problems without adding the trigger Hamiltonian to the Hamiltonian~(\ref{eq_annealing}). The lines are fits to the data.}
    \label{fig:12_nonstoq_orig}
\end{figure}

\begin{figure*}[ht]
    \centering
     \begin{minipage}[c]{0.32\textwidth}
         \includegraphics[scale=0.4]{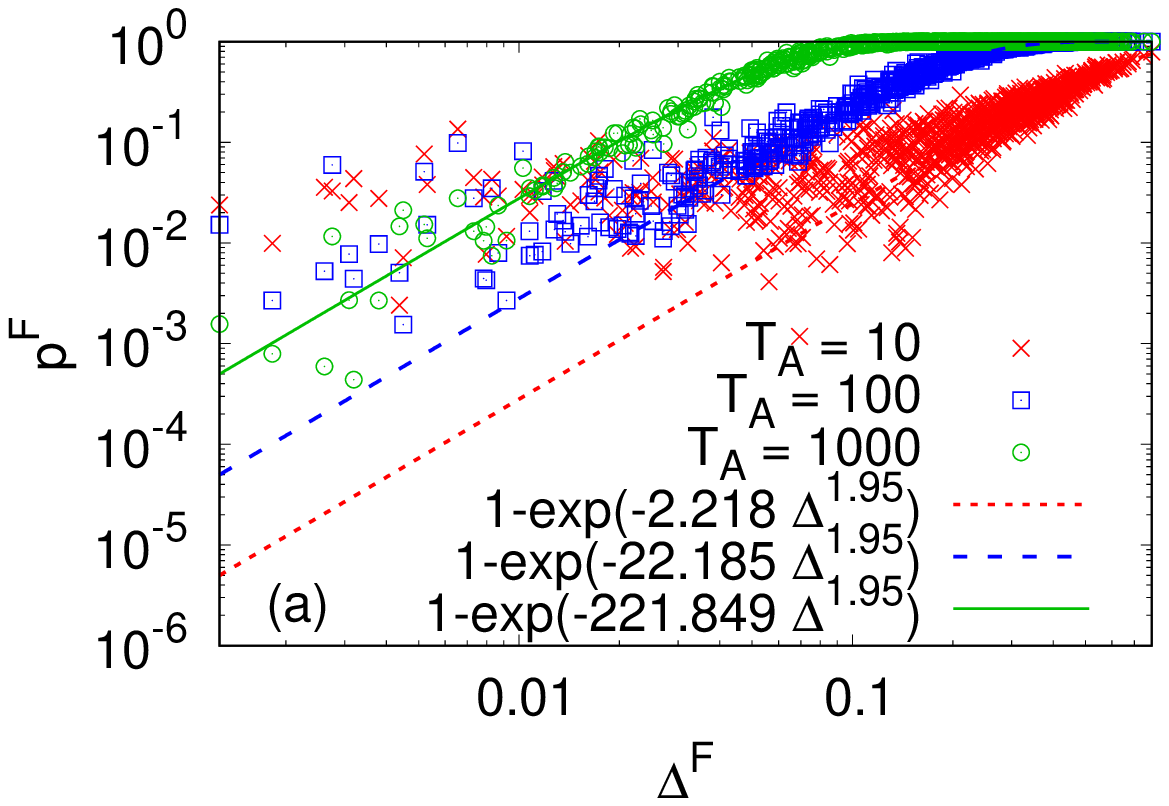}
     \end{minipage}
     \begin{minipage}[r]{0.3\textwidth}
         \includegraphics[scale=0.4]{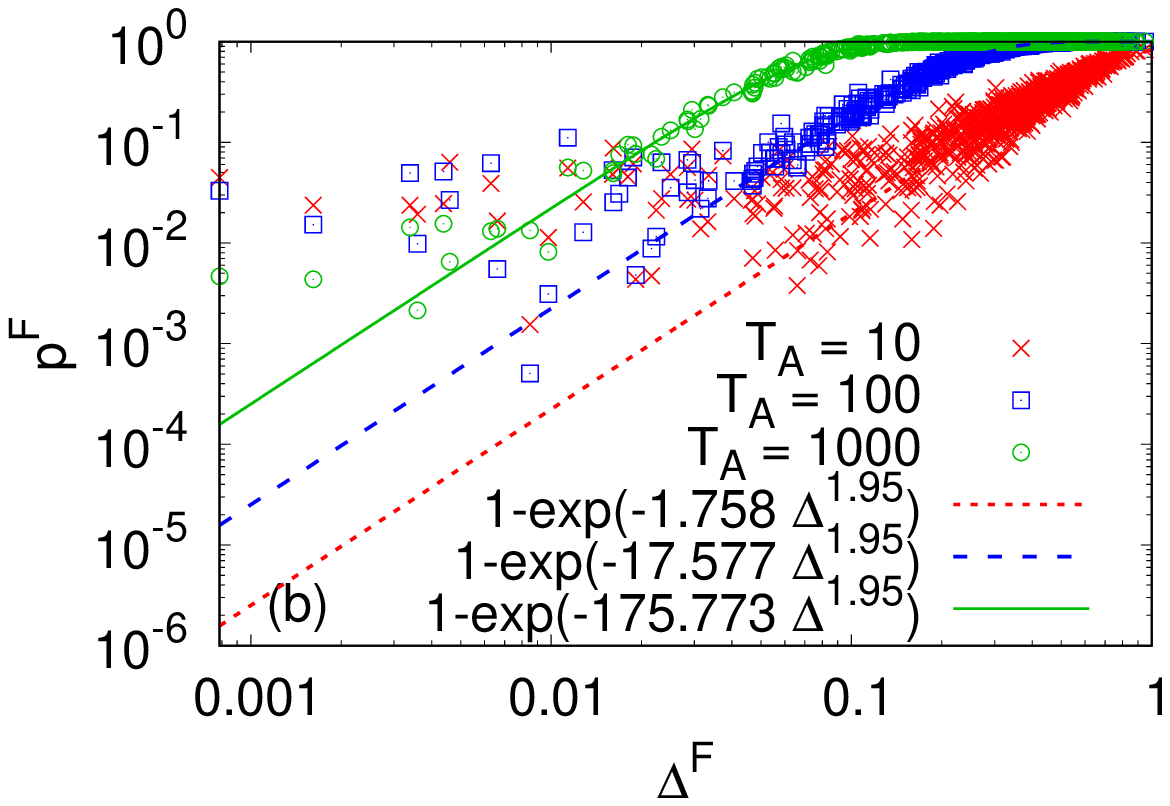}
     \end{minipage}
     \begin{minipage}[r]{0.3\textwidth}
         \includegraphics[scale=0.4]{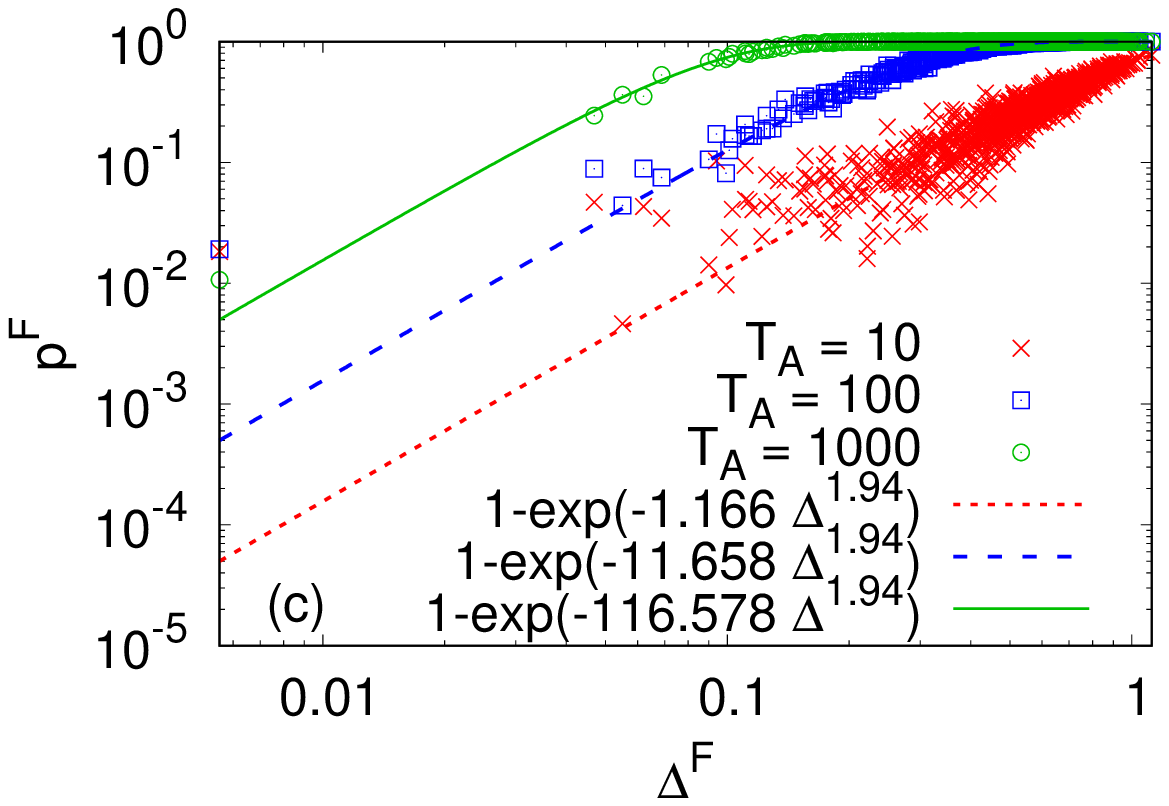}
     \end{minipage}\\
     \begin{minipage}[c]{0.32\textwidth}
         \includegraphics[scale=0.4]{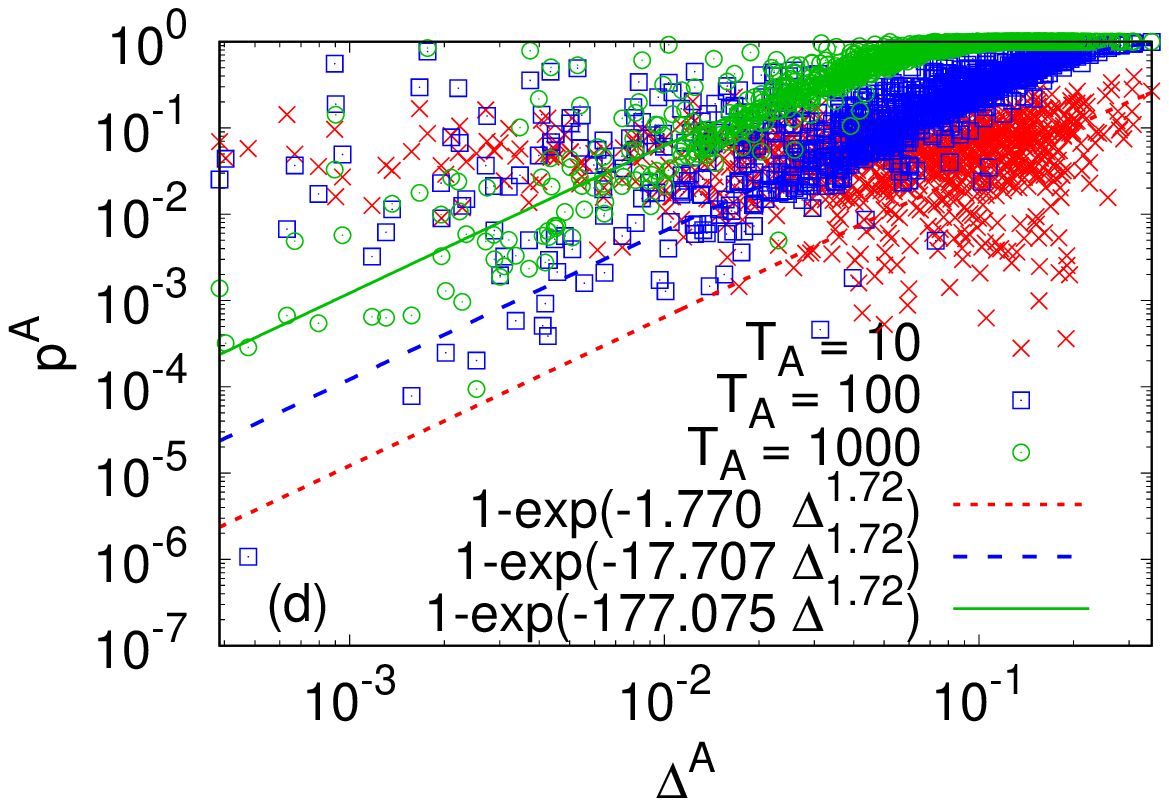}
     \end{minipage}
     \begin{minipage}[r]{0.3\textwidth}
         \includegraphics[scale=0.4]{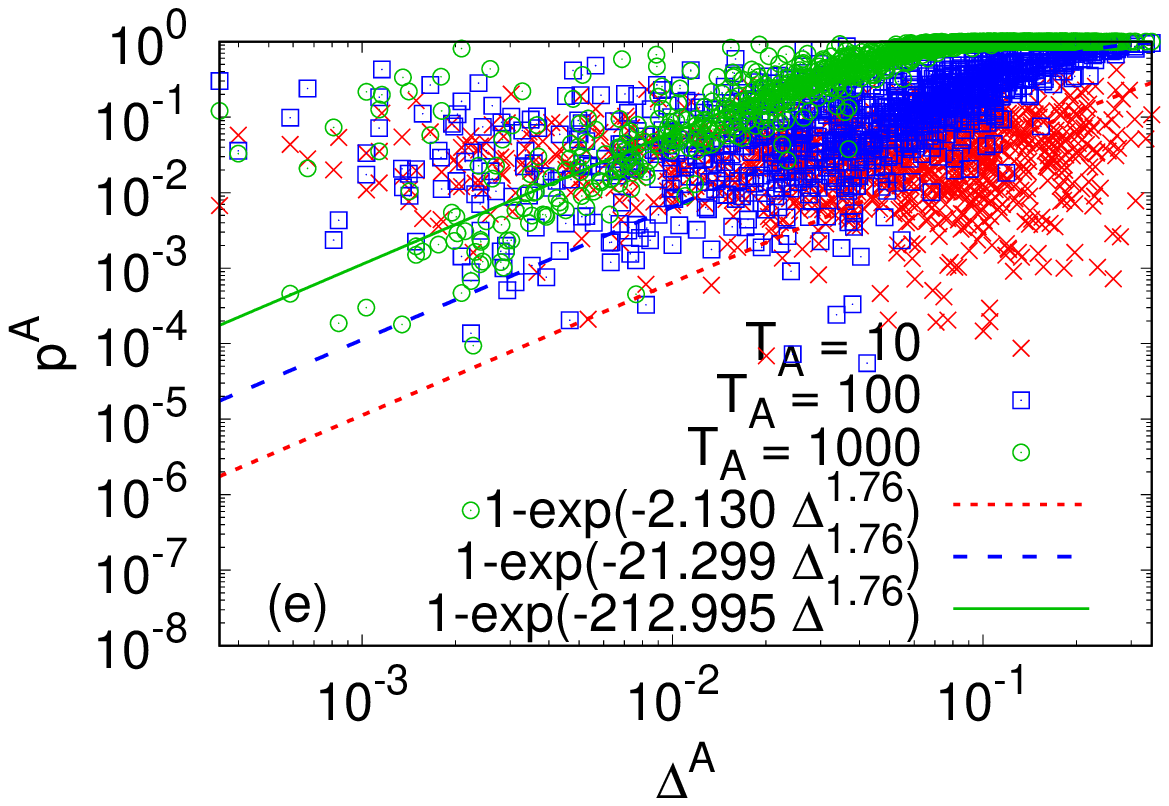}
     \end{minipage}
     \begin{minipage}[r]{0.3\textwidth}
         \includegraphics[scale=0.4]{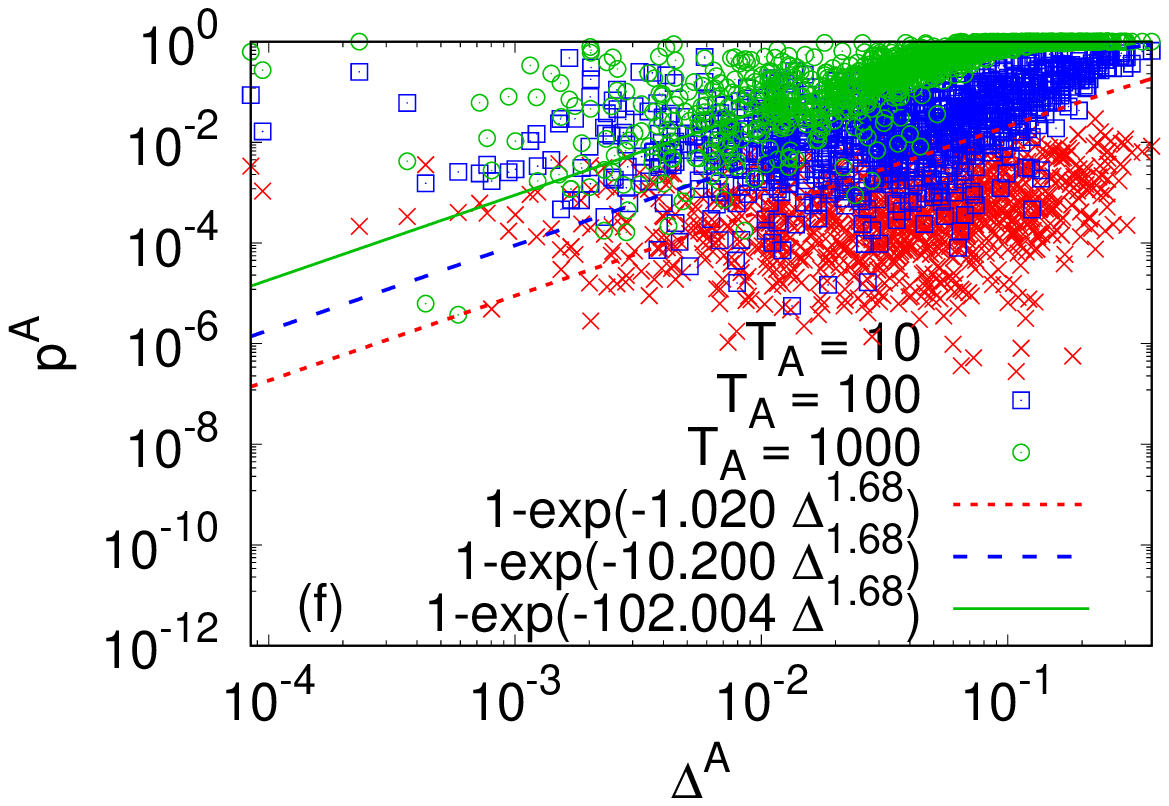}
     \end{minipage}\\
    \caption{(Color online) Success probability $p$ versus minimum energy gap  $\Delta$ for 12-variable nonstoquastic problems after adding the (a)-(c) ferromagnetic and (d,)-(f) antiferromagnetic trigger Hamiltonian to the Hamiltonian~(\ref{eq_annealing}), with trigger strengths (a), (d) $g$=0.5, (b), (e) $g$=1.0, and (c), (f) $g$=2.0. The lines are fits to the data.}
    \label{fig:nontoq_12_succvsgap}
\end{figure*}

For completeness, we show the results obtained for the analysis of the 12-variable nonstoquastic problems. Figure~\ref{fig:12_nonstoq_orig} shows the plot for the success probability versus minimum energy gap for the nonstoquastic problems without adding the trigger Hamiltonian to Hamiltonian~(\ref{eq_annealing}). Figure~\ref{fig:nontoq_12_succvsgap} shows the same upon adding the trigger Hamiltonians.

\bibliography{reference}

\end{document}